\documentclass[letterpaper,twocolumn,10pt]{article}
\usepackage{usenix2019_v3}

\usepackage{tikz}
\usepackage{tabularx}
\usepackage{amsmath}
\usepackage{amssymb}
\usepackage{diagbox}
\usepackage{pifont}
\usepackage{adjustbox}
\usepackage{multirow}
\usepackage{caption}
\usepackage{subfig}
\newcommand{\cmark}{\ding{51}}%
\newcommand{\xmark}{\ding{55}}%
\usepackage{filecontents}

\usepackage[subtle]{savetrees}

\usepackage[nolist,nohyperlinks]{acronym}
\usepackage{mathtools}
\usepackage[inkscapeformat=png]{svg}

\newcommand{\Mod}[1]{\ \mathrm{mod}\ #1}
\usepackage{siunitx}
\DeclareSIUnit{\GHz}{GHz}
\DeclareSIUnit{\MHz}{MHz}
\DeclareSIUnit{\kHz}{kHz}
\DeclareSIUnit{\ppm}{ppm}
\DeclareSIUnit{\bits}{bits}
\usepackage{paralist}  
\usepackage{flushend}

\begin{acronym}
\acro{adc}[ADC]{Analog-to-Digital Converter}
\acro{ber}[BER]{Bit-Error-Rate}
\acro{bprf}[BPRF]{Base Pulse-Repetition Frequency}
\acro{cfo}[CFO]{Clock Frequency Offset}
\acro{cir}[CIR]{Channel Impulse Response}
\acro{dsp}[DSP]{digital signal processing}
\acro{edlc}[ED/LC]{Early-Detect/Late-Commit}
\acro{currentstd}[802.15.4z]{802.15.4z}
\acro{dac}[DAC]{Digital-to-Analog Converter}
\acro{hrp}[HRP]{High Rate Pulse Repetition Frequency}
\acro{lrp}[LRP]{Low Rate Pulse Repetition Frequency}
\acro{v1}[MD]{Mix-Down}
\acro{v2}[S\&A]{Stretch-and-Advance}
\acro{mmsuwb}[NBA-MMS-UWB]{Narrowband-Assisted Multi-Millisecond Ultra-Wide Band}
\acro{nb}[NB]{Narrowband}
\acro{newstd}[802.15.4ab]{802.15.4ab}
\acro{dstwr}[DS-TWR]{Double-Sided Two-Way Ranging}
\acro{pkes}[PKES]{Passive Key-Less Entry and Start}
\acro{pre}[RSF]{Ranging Sequence Fragment}
\acro{prf}[PRF]{Pulse Repetition Frequency}
\acro{rif}[RIF]{Ranging Integrity Fragment}
\acro{rsf}[RSF]{Ranging Sequence Fragment}
\acro{rrc}[RRC]{Root-Raised Cosine}
\acro{rssi}[RSSI]{Received Signal Strength Indication}
\acro{sfo}[SFO]{Sampling Frequency Offset}
\acro{snr}[SNR]{Signal-to-Noise Ratio}
\acro{sstwr}[SS-TWR]{Single-Sided Two-Way Ranging}
\acro{sts}[STS]{Scrambled Timestamp Sequence}
\acro{tdoa}[TDoA]{Time Difference of Arrival}
\acro{toa}[ToA]{Time of Arrival}
\acro{tof}[ToF]{Time of Flight}
\acro{uwb}[UWB]{Ultra-Wide Band}
\acro{sdk}[SDK]{Software Development Kit}
\acro{rtls}[RTLS]{Real Time Location Service}
\acro{sdr}[SDR]{Software Defined Radio}
\acro{nb}[NB]{Narrow Band}
\end{acronym}

\newcommand{\aref}[1]{\hyperref[#1]{Appendix~\ref{#1}}}

\begin{document}

\date{}

\title{\Large \bf Time for Change: How Clocks Break UWB Secure Ranging}

\author{
{\rm Claudio Anliker}\\
ETH Zurich
\and
{\rm Giovanni Camurati}\\
ETH Zurich
 \and
 {\rm Srdjan  \v{C}apkun}\\
ETH Zurich
} 

\maketitle


\begin{abstract}
Due to its suitability for wireless ranging, Ultra-Wide Band (UWB) has gained traction over the past years. UWB chips have been integrated into consumer electronics and considered for security-relevant use cases, such as access control or contactless payments. However, several publications in the recent past have shown that it is difficult to protect the integrity of distance measurements on the physical layer. In this paper, we identify transceiver clock imperfections as a new, important parameter that has been widely ignored so far. We present \textit{Mix-Down} and \textit{Stretch-and-Advance}, two novel attacks against the current (IEEE 802.15.4z) and the upcoming (IEEE 802.15.4ab) UWB standard, respectively.  
We demonstrate \textit{Mix-Down} on commercial chips and achieve distance reduction from \SI{10}{\meter} to \SI{0}{\meter}. For the \textit{Stretch-and-Advance} attack, we show analytically that the current proposal of IEEE 802.15.4ab allows reductions of over \SI{90}{\meter}. In order to prevent the attack, we propose and analyze an effective countermeasure.
\end{abstract}

\section{Introduction}
UWB has experienced a revival over the the last years as one of the main technologies for short-distance wireless ranging. Because of its wide bands, it can achieve sub-decimeter accuracy. UWB ranging computes the distance between two devices based on the Time of Flight (ToF) of exchanged messages. This makes it inherently secure against \emph{simple} relay attacks, which were conducted successfully against Received Signal Strength Indication (RSSI)-based ranging schemes, like those used in Passive Keyless Entry and Start (PKES) systems~\cite{DBLP:conf/ndss/FrancillonDC11}. 

Due to these attacks, the standardization of UWB ranging progressed at a high pace, and several car vendors have already integrated UWB chips into their vehicles and keys \cite{bmw-uwb, jaguar, hyundai, ccc}. Companies like Apple, Samsung, Google, and Xiaomi have released smart phones, tags, and other gadgets with embedded UWB transceivers~\cite{samsung-smarttag, apple-airtag-ranging, samsung-gv60, pixel6, xiaomi}. BMW and Apple announced a collaboration enabling customers to unlock their car using an UWB-equipped iPhone, indicating that UWB is going to be used in this domain in the near future~\cite{bmw-uwb}. However, protecting the integrity of distance measurements on the physical layer has proven to be difficult: attacks such as Cicada~\cite{DBLP:journals/twc/PoturalskiFPHB12,5616900}, Early-Detect/Late-Commit~\cite{DBLP:journals/twc/PoturalskiFPHB11} and, most recently, Ghost Peak~\cite{leu2022ghost} show that distance reductions are still possible. 

In this paper, we introduce a new class of attacks on UWB ranging. By manipulating UWB signals over the air, we attack the chips' tolerance towards clock errors. We analyze how UWB device clocks are used and synchronized, and present two attacks on UWB ranging standards and implementations, which we call Mix-Down (MD) and Stretch-and-Advance (S\&A). Our attacks exploit drifts and subsequent synchronizations of transceiver clocks as a novel attack vector, which has not been discussed in the context of ranging integrity so far. Our contributions can be summarized as follows:

\textbf{Discussion of the security implications of clocks:} 
We explain the functional consequences of non-ideal transceiver clocks, namely the carrier frequency offset (CFO) and the sampling frequency offset (SFO), and show how they are mitigated in practice. We discuss how these mitigations can be abused to manipulate distance measurements over the air. 

\textbf{Two novel attacks against UWB ranging:} We present Mix-Down (MD), an attack against Single-Sided Two-Way
Ranging (SS-TWR), and Stretch-and-Advance (S\&A), which affects both SS-TWR and Double-Sided Two-Way Ranging (DS-TWR). Both are distance reduction attacks that heavily depend on design parameters, such as clock accuracy and message length. We show that unfavourable parameter choices can lead to distance reductions of up to dozens of meters. Concretely, we show that MD can be exploited in chips following the current standard. While the practical impact of S\&A is currently negligible, the attack may lead to considerable reductions in the future standard 802.15.4ab. \autoref{tab:attack_comparison} provides an overview of the standards and modes affected by either attack.

\textbf{Practical evaluation of MD:} We conduct experiments to demonstrate how MD can be used to attack ranging chips that use SS-TWR and comply with the current IEEE 802.15.4z standard, showing that even conservative parameters allow distance reductions of several meters. We show that SS-TWR with real-time compensation is fundamentally insecure and should not be used in security-sensitive applications.

\textbf{Countermeasure against Stretch-and-Advance:} We introduce a countermeasure to secure DS-TWR against our attacks. We recommend that this proposal be considered in the upcoming IEEE 802.15.4ab standard.

\textbf{Responsible disclosure:} We contacted leading chip vendors and informed them about both attacks prior to publication.

\begin{table}
\centering
\begin{tabular}{l|cccc}
\hline
Attack & 4z SS & 4z DS & 4ab SS & 4ab DS\\
 \hline\hline
 MD & \cmark & \xmark & \textasciitilde\cmark & \xmark \\ 
 S\&A & \xmark* & \xmark* & \textasciitilde\cmark & \textasciitilde\cmark \\    
 \hline
\end{tabular}
\caption{Overview on standards / ranging modes affected by each attack. Notation: "\cmark" means affected and experimentally verified, "\textasciitilde\cmark" potentially affected future standard, "\xmark*" theoretically affected but negligible attack margin, "\xmark" not affected.}
\label{tab:attack_comparison}
\end{table}

\section{Background: UWB Ranging}\label{sec:uwb}

This section provides an overview over ranging with UWB and discusses effects related to errors of transceiver clocks that are relevant for UWB security.

\subsection{UWB Ranging Standards}

The UWB channels relevant for this paper are roughly \SI{500}{MHz} wide. This enables UWB transceivers to produce radio signals with very short pulses (\textit{i.e.} \SI{2}{ns}). The Time of Arrival (ToA) of a signal is easier to measure with such pulses than, for example, with WiFi or BlueTooth waveforms. This makes UWB an attractive candidate for wireless ranging. The current standard regulating UWB ranging is IEEE 802.15.4z, published in 2020~\cite{9144691,9179124}. A new version, 802.15.4ab, is currently under development~\cite{abdocs}.

\subsubsection{Single-Sided and Double-Sided Two-Way Ranging}
\begin{figure}[]
    \includegraphics[width=\linewidth]{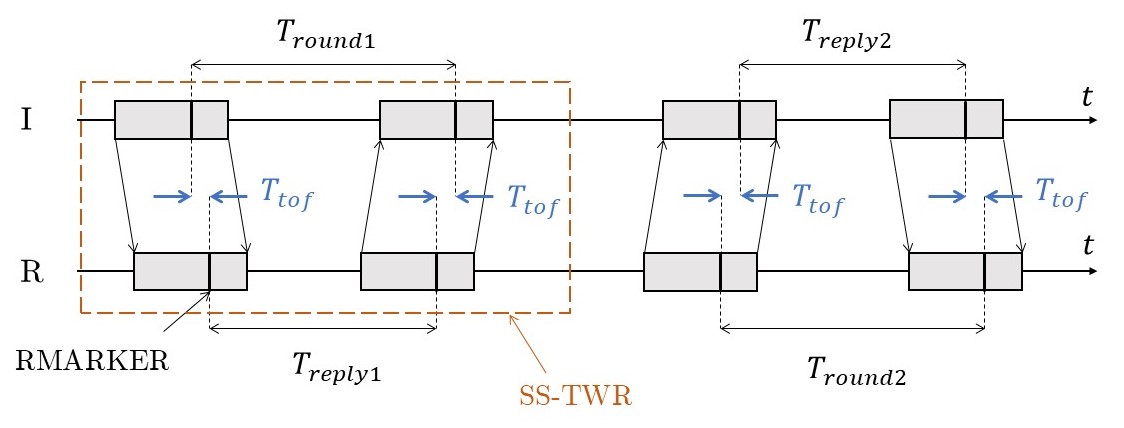}
    \caption{SS-TWR (dashed box) and DS-TWR between the initiator and responder in 802.15.4z. }
    \label{fig:sstwr_dstwr}
\end{figure}

On the protocol layer, the current UWB standard supports three modes for ranging, namely SS-TWR, DS-TWR, and ranging based on  Time Difference of Arrival (TDoA). In this paper, we focus on SS-TWR and DS-TWR, which are both shown in \autoref{fig:sstwr_dstwr}. As the names suggest, both modes require a back-and-forth or two-way exchange of ranging messages to estimate the distance between the \textit{initiator}, who sends the first ranging message, and the \textit{responder}. We will stick to these names, whenever their roles are relevant, to be consistent with the nomenclature of the 802.15.4z standard. When we use the terminology \textit{transmitter} and \textit{receiver}, we simply indicate in which direction a message is being sent, regardless of the device's role in SS-TWR/DS-TWR.

The ToA of a UWB packet is measured at a predefined position of the packet, which is called the RMARKER. Since this is just a point of reference, it can be put in any position, as long as it is used consistently. In 802.15.4z, the RMARKER is at the end of the preamble, \textit{i.e.}, roughly in the middle of the packet, as depicted in \autoref{fig:sstwr_dstwr}.

\textbf{SS-TWR}: In the simpler case of SS-TWR, only two UWB messages are sent. The distance between the devices is then estimated based on the ToF of these messages, which can be computed by subtracting the known processing time from the entire duration of the exchange:
\begin{equation}
T_{ToF} = \frac{1}{2}\cdot(T_{round1} - T_{reply1}(1-c))
\end{equation}
Note that there is a correction term $c$ related to the clock drift between the two devices. This difference in clock speed means that, from the initiator's perspective, the responder's clock is either too fast or too slow. Therefore, it sent the reply either too early or too late. The reply time deducted by the initiator would be incorrect in this case, and the ToF calculation would be inaccurate. The initiator compensates for this clock drift and rectifies the ranging measurement by scaling $T_{reply}$. As a matter of fact, this correction is the enabler of the Mix-Down attack we present in \autoref{sec:v1}.

\textbf{DS-TWR}: In this mode, the transceivers preform two complete exchanges, one in each direction. The ToF can be computed as:
\begin{equation}\label{eq:dstwr}
T_{ToF} = \frac{T_{round1} \cdot T_{round2} - T_{reply1} \cdot T_{reply2}}{T_{reply1} + T_{reply2} + T_{round1} + T_{round2}}
\end{equation}

This formula implicitly corrects errors introduced by the aforementioned clock drift, but it forces the responder to send $T_{round2}$ to the initiator. The reason why this formula is used (instead of taking the average of the two rounds) is that it works well in cases where the reply times of the two devices are not symmetric~\cite{DBLP:conf/wpnc/NeirynckLM16}. The formula also averages out differences in the ToF of the messages. Generally speaking, the DS-TWR mode requires longer media access and has a higher latency due to the additional messages it sends. On the other hand, it is less prone to inaccuracies induced by the clock drift between devices. 

We have looked at an assortment of gadgets in the Apple and Samsung ecosystems, such as the iPhone 11 Pro, the AirTag, the HomePod, the Samsung Galaxy S10, and the Samsung Smart Tag. By default, all of them use DS-TWR. We could not find any official statements as to why this is the case. We assume that the motivation is the ToF computation, whose accuracy does not rely on a clock drift estimation. Furthermore, the additional delay and media access incurred by using this mode do not seem to be a hindrance in existing use cases. 
Other products, such as the Decawave Positioning and Networking Stack R2 in  MDEK1001~\cite{mdek1001}, use SS-TWR, as its lower power consumption and latency are advantageous in Real Time Location Service (RTLS ) systems. Furthermore, several open-source projects~\cite{opensource1,opensource2,opensource3} employ SS-TWR, and SS-TWR is considered in the context of one-to-many ranging in the upcoming standard \cite{1tomany}. The Software API Guide of the Qorvo DWM3000~\cite{DWM3000EVBQorvo} (v2.2, 6.3.26) mentions the newly available clock drift measurement in the driver as a feature that makes SS-TWR more accurate, and explicitly states that it does no longer recommend DS-TWR over SS-TWR for reasons of accuracy. To the best of our knowledge, there is no indication in the publicly accessible documentation of any vendor that one mode should be preferred over the other based on security arguments.

\subsubsection{High and Low Pulse Repetition Frequency modes}
UWB generally does not have exclusive access to the frequency bands it uses and is subject to stringent power regulations. This is a challenge for UWB transceivers, as it limits the link budget of ranging messages. 802.15.4z comprises two physical layer designs that approach the power limits differently. First, the High Rate Pulse Repetition Frequency (HRP) mode uses a high number of narrowly spaced pulses, each of which has to be send at low power. On the other hand, the Low Rate Pulse Repetition Frequency (LRP) mode opts for higher peak power at the cost of sending fewer, more widely spaced pulses. This has lead to fundamentally different transceiver designs. In the context of this paper, the differences between LRP and HRP are largely irrelevant.

\subsubsection{802.15.4ab: Future of UWB ranging}\label{sec:mms_uwb}
A major drawback of the current UWB standard are the mentioned link budget limitations. UWB chips have found adoption in a lot of small gadgets, such as mobile phones, tags, and smart watches that do not have high-gain antennas at their disposal~\cite{apple-uwb, samsung-gv60, apple-airtag-ranging, samsung-smarttag}. Consequently, increasing the link budget is an important goal of the upcoming 802.15.4ab standard. To that end, Apple proposed Narrowband-Assisted Multi-Millisecond Ultra-Wide Band (NBA-MMS-UWB) ranging~\cite{mmsuwb-apple-first}. The main idea behind the proposal is to use long ranging messages that comprise several fragments at one-millisecond intervals. This increases the signal's energy and, therefore, the link budget. Since the fragments are succeeded by intervals of no transmission, the mean Power Spectral Density does not exceed the limits imposed on UWB communication. However, the multi-millisecond ranging messages also have a drawback: they are more susceptible to clock-related errors, which we discuss in detail in \autoref{sec:clocks}.

\begin{figure}[t]
\includegraphics[width=\linewidth]{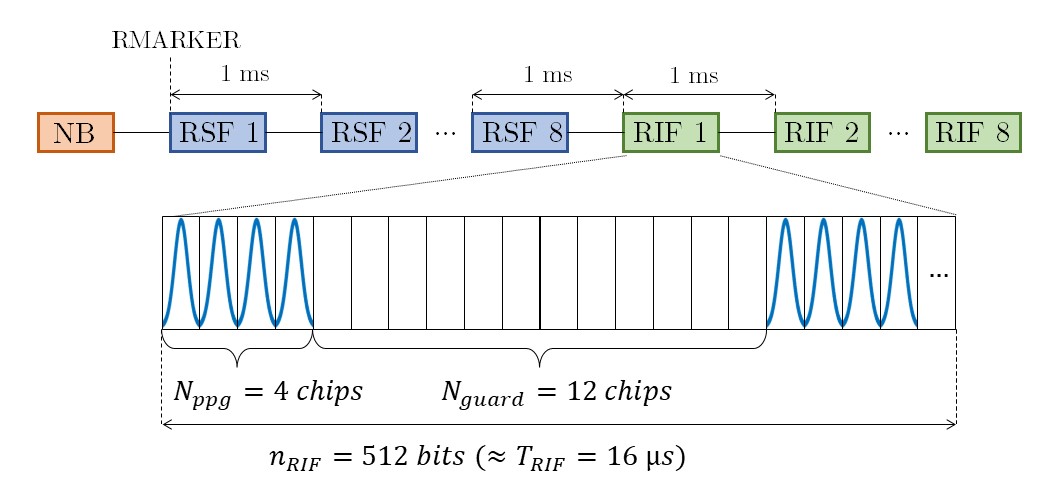}

\caption{The proposed NBA-MMS-UWB frame consists of a narrowband (NB) fragment followed by Ranging Sequence Fragments (RSFs) for measuring the ToA (RMARKER), and Ranging Integrity Fragments (RIFs) for ToA integrity verification. Following~\cite{david}, each RIF contains $n_{RIF}$ decodable bits. Using multiple fragments (\textit{e.g.}, here $N_{RSF}=N_{RIF}=8$) increases the link budget at the cost of a longer transmission time.} 
\label{fig:3db_rif}
\end{figure}

NBA-MMS-UWB works as follows: before the UWB ranging starts, narrow-band messages (NB) are exchanged between the devices to transmit necessary data. The UWB part consists of Ranging Sequence Fragments (RSFs) to establish the ToA and Ranging Integrity Fragments (RIFs), whose purpose is to prevent distance reduction attacks. The RMARKER is placed at the beginning of the first RSF~\cite{nbauwb-draft0}. An example configuration is shown in \autoref{fig:3db_rif}. The standard envisions frames with between 1 and 8 RSFs and RIFs. At the time of writing, the implementation of the RIF was still a matter of ongoing discussion, and two alternatives had been submitted to the task group.

The first proposal retains the Scrambled Timestamp Sequence (STS) waveform from 802.15.4z HRP~\cite{nbauwb-draft0}. This would imply a ToA verification based on an approximated Channel Impulse Response (CIR), which can be obtained by cross-correlating the incoming signal with a local STS template. Alternatively, the STS CIR could be used to establish and verify the ToA at the same time. 

The second proposal from Barras et. al \cite{david} suggests modulating the pseudo-random bits as bursts of $N_{ppg}\in \{1,2,4,8\} $ pulses with a peak Pulse Repetition Frequency (PRF) of 499.2\si{MHz}, which corresponds to one chip ($\approx$ 2\si{ns}) per pulse. The bursts are separated by guard intervals that reduce the mean PRF to 124.8\si{MHz}. Consequently, their duration depends on $N_{ppg}$ and is three times the one of a burst. For $N_{pgg}=1$, the scheme coincides with the legacy STS waveform mentioned before. However, according to the authors, choosing $N_{ppg}\in \{4,8\}$ leads to guard intervals long enough to sufficiently reduce inter-burst interference. This should enable receivers to decode and verify bits individually.  

From a security perspective, this is a major advantage over the first, CIR-based approach, whose security guarantees depend heavily on receiver specifics and are neither well understood nor easy to prove\cite{leu2022ghost}.

Therefore, we assume throughout this work that the second proposal is used. Where we provide concrete examples, we choose the configuration shown in \autoref{fig:3db_rif}. With $N_{ppg} = 4$ pulses per burst/bit, it strikes a balance between inter-burst interference and symbol length. This compromise simplifies decoding of individual bits, while limiting the attack window for Early-Detect/Late-Commit~\cite{DBLP:journals/twc/PoturalskiFPHB11}. This results in $T_{bit}=32\si{ns}$ (16 chips) per bit/burst, of which 8\si{ns} (4 chips) are active. We further assume that a RIF is $N_{s}=32$ units of 512 chips long. One RIF contains $n_{RIF}= 512$ bits, and the entire ranging message comprises $N_{RIF}=8$ RIFs, resulting in $n_{tot}=4096$ bits in total.

\subsection{Clocks}\label{sec:clocks}
A digital transceiver uses a clock reference derived from a crystal oscillator to measure time intervals, sample incoming signals, and generate radio frequencies.
Because of oscillator imperfections and environmental factors, the clock frequency of a device and all derived quantities differ from their nominal values.
For example, an initiator with a nominal clock frequency $f_{nom}$ would actually run at a different clock frequency $f_{init}$.
This error is conveniently expressed in parts per million (\si{ppm}) and called the \textit{clock drift}\footnote{We refrain from using the term \textit{clock frequency offset} to avoid confusion with the Carrier Frequency Offset (CFO), which is caused by the clock drift.} $c_{init}$  (\si{ppm}):
\begin{equation}\label{eq:clockdrift}
    c_{init}[\si{ppm}]=\frac{f_{init}-f_{nom}}{f_{nom}}
\end{equation}
The UWB standard~\cite{9179124} defines the accepted range of clock drifts as $\pm20\si{ppm}$. 
While clock drifts might vary over time, they can be considered to be constant during the short duration of a ranging exchange.

A device with a negative clock drift will generate a carrier frequency that is lower than its nominal value. If asked to wait for a nominal time interval, it will wait for a longer time. If asked to measure an absolute time interval between two events, it will report a smaller value. The opposite will occur for a device with a positive clock drift.
The larger the times and frequencies involved, the larger the absolute errors in \si{\second} and \si{\hertz} caused by the same clock drift in \si{\ppm}.

In general, transmitter and receiver exhibit different clock drifts, causing the following problems. First, the carrier frequencies they generate differ by a Carrier Frequency Offset (CFO) expressed in \si{\hertz}.
The CFO causes a rotation of the symbols on the constellation plane, which could lead to decoding errors.
Second, their sampling frequencies are affected by a Sampling Frequency Offset (SFO), causing the receiver to sample the incoming symbols at the wrong intervals. As a result, the receiver accumulates less energy per symbol, which degrades performance.
Finally, transceivers have different estimates of time intervals, which impacts the accuracy of distance measurements based on the ToF. 

Upon reception of a new message from the transmitter, the receiver can derive the clock drift between the two from the CFO~\cite{DBLP:journals/sensors/SidorenkoSSAH19}, and apply the necessary corrections. 



\section{The Mix-Down Attack (MD)}\label{sec:v1}

In this section, we describe the Mix-Down (MD) attack, the first of two novel attacks related to device clocks in UWB ranging. The underlying vulnerability affects the SS-TWR mode of the current standard (802.15.4z) and of the current proposal of the upcoming standard (802.15.4ab).

\subsection{Threat Model}
\label{subsec:threat-model}
The goal of the attacker is to reduce the distance measured by two UWB transceivers. 
We assume a Dolev-Yao attacker that controls the wireless channel and can manipulate, block, inject, and eavesdrop all wireless signals. Furthermore, the attacker is able to precisely configure power and timing of their transmission if required. Following Kerckhoff's principle, we we also assume that the attacker knows everything about the victim devices except their cryptographic secrets.

This threat model decouples the security of the system from factors related to specific scenarios or use cases, such as the spatial arrangement of the devices, and from obscure implementation details, which cannot be openly analyzed or verified. A secure ranging scheme should provide a high and verifiable level of protection against distance reduction attacks under this strong threat model.

In practice, attacks on wireless ranging succeed under far less ideal conditions: many instances of car thefts have shown that parked vehicles are a viable target, even if the exact position of they key fob was not known.\footnote{\url{https://www.youtube.com/watch?v=PEMOWPj2i-0}}\textsuperscript{,}\footnote{\url{https://www.youtube.com/watch?v=bR8RrmEizVg}}\textsuperscript{,}\footnote{\url{https://www.youtube.com/watch?v=bR8RrmEizVg}.}

\subsection{General Idea}\label{sec:v1_generalidea}
Single-Sided Two-Way Ranging (SS-TWR) is depicted in \autoref{fig:sstwr_dstwr}. The initiator computes the ToF by measuring the round-trip time of two messages with the responder, and subtracting the reply time:

\begin{equation}\label{eq:sstwr}
 T_{ToF} = \frac{1}{2}\Big(T_{round} - T_{reply}(1 - c)\Big)
\end{equation}

The corrective term $c = c_{resp} - c_{init}$ measured by the initiator denotes the \textit{relative} drift between the clocks of the two devices. It is the difference of the \textit{absolute} clock drifts (\textit{i.e.}, with reference to an ideal clock) of the responder and the initiator, respectively. Due to these drifts, the transceivers are in disagreement about how long, for example, \SI{1}{ms} or $T_{reply}$ last in absolute time. 

The formula above implies that the SS-TWR initiator assumes its own clock to be correct. From the initiator's perspective, a relative clock drift $c$ means that the responder clock is imprecise, and that the reply time it waited is therefore incorrect. Since a deviation of a single nanosecond results in a ranging error of $\pm \SI{15}{cm}$ in \autoref{eq:sstwr}, the initiator must estimate the relative clock drift and compensate accordingly. As described in \autoref{sec:clocks}, clock drift estimation can be done by measuring the Clock Frequency Offset (CFO) of the incoming signal.

In the MD attack, we exploit that this estimation can be controlled by the attacker. We achieve this by replaying the response message on a different carrier frequency that corresponds to a strong negative clock drift $c_{att}$. The initiator receives the replayed message and interprets $c_{att}$ as $c_{resp}$. It concludes that the responder's clock is too slow, and, consequently, that it must have replied to late. To compensate for this apparent delay, the initiator deducts it from what would be the correct ToF. The final, shorter ToF results in a distance reduction. 

This attack only changes the frequency of the signal carrying the message, but not the content of the message itself. Therefore, it cannot be prevented by existing security mechanisms relying on the pseudo-randomness of bits/pulses, such as the STS in 802.15.4z HRP. Furthermore, the MD attack is fully deterministic and theoretically works on every ranging round. Since the attacker can choose $c_{att}$, they can also steer the distance reduction within certain limits. This is in stark contrast to attacks like Ghost Peak~\cite{leu2022ghost}, which are probabilistic and whose measurement outcomes are beyond the attacker's control. Finally, the MD attack could also be used to increase the ToF by faking a faster responder clock. We do not discuss this in detail, since distance enlargement is usually trivial to achieve by means of delayed retransmission.

\subsection{Analysis}\label{sec:v1_analysis}

We can express the result of the attack as  
\begin{equation}\label{eq:sstwr_attack}
\begin{aligned}
T_{ToF} + \delta_{ToF} &= \frac{1}{2}\Big(T_{total} - T_{reply}(1 - c')\Big)\\
\end{aligned}
\end{equation}
where  $\delta_{ToF}$ denotes the ToF difference caused by the attack. Note that the attacker's interference changed $c=c_{resp}-c_{init}$ to $c'=c_{att}-c_{init}$, so that the initiator measures the clock drift between its own clock and the attacker's. It follows from this equation that a ToF reduction is a negative quantity, which is the convention we will follow throughout this paper. For simplicity, and because they are proportional, we will treat the terms "ToF reduction" and "distance reduction" synonymously. 
Subtracting \autoref{eq:sstwr}  from \autoref{eq:sstwr_attack} results in

\begin{equation}\label{eq:deltatof}
\begin{aligned}
 \delta_{ToF} &= \frac{T_{reply}}{2}\cdot \delta_{c}
\end{aligned}
\end{equation}
where
\begin{equation}\label{eq:deltac}
    \begin{aligned}
    \delta_{c} = c'-c = c_{att} - c_{resp}\\
    \end{aligned}
\end{equation}
The value of $\delta_{ToF}$ therefore depends on the length of the reply time as well as on the difference $\delta_{c} = c_{att}-c_{resp}$ between the clock drift chosen by the attacker and the genuine clock drift of the responder. To cause a distance reduction ($\delta_{c}<0$) the attacker must use a clock drift $c_{att}$ less than the clock drift $c_{resp}$ of the responder. The maximal distance reduction is then bounded by the range of clock drifts accepted by the initiator.

An acceptable range for $c_{init}$ and $ c_{resp}$ is $[-c_{std}, c_{std}]$, where $c_{std}$ is defined in the UWB standard (\textit{e.g.}, $c_{std}=\SI{20}{ppm}$ in 802.15.4z). For simplicity, we first consider an initiator who has an ideal clock ($c_{init}=\SI{0}{ppm}$) and is aware of its precision. The latter enables the initiator to limit $c = c_{resp} \in[-\SI{20}{ppm}, +\SI{20}{ppm}]$, which covers all standard-compliant responder clocks. The attacker is also limited to $c_{att} \in[-\SI{20}{ppm}, +\SI{20}{ppm}]$ under these conditions. They are generally interested in the highest negative $c_{att}$, as this causes the largest reduction.
In the best case for the attacker, with $c_{resp}=+\SI{20}{\ppm}$ and $c_{att}=-\SI{20}{\ppm}$, we have $\delta_{c}=-\SI{40}{\ppm}$. For an assumed reply time of $T_{reply}=\SI{2}{\milli\second}$, this results in a distance reduction of -\SI{12}{\meter}.

Transceiver clocks are not ideal in general, since hardware imperfections and environmental factors cause them to diverge. In the worst case, initiator and responder could \textit{unknowingly} be affected by any $c_{init},c_{resp}\in[-\SI{20}{ppm}, +\SI{20}{ppm}]$. This would force the initiator to accept incoming signals with $c \in [-2c_{std}, 2c_{std}]$. 
Under this condition, the attacker may choose $c_{att}\in [c_{init}-2c_{std}, c_{init} + 2c_{std}]$.
The best-case scenario for the attacker is  $c_{init}=-\SI{20}{\ppm}$ and $c_{resp}=+\SI{20}{\ppm}$. Choosing $c_{att}=-\SI{60}{\ppm}$ leads to $\delta_{c}=-\SI{80}{\ppm}$, and, for $T_{reply}=\SI{2}{\milli\second}$, to a distance reduction of -\SI{24}{\meter}. In the average case of $c_{init}=c_{resp}=0\si{ppm}$, the reduction limit is -\SI{12}{\meter}.
\subsection{Experimental Evaluation}\label{sec:v1_evaluation}

In this section, we demonstrate distance reductions against commercial HRP UWB devices in an indoor environment.
The MD attacker is build of simple off-the-shelf components and two software-defined radios. It is capable of measuring the clock drift between any two devices over-the-air, which we leverage to tune its own clock drift before an attack.

\subsubsection{Setup and Practical Aspects}
\textbf{Test devices:}
We use two Qorvo DWM3000EVB boards~\cite{DWM3000EVBQorvo} as initiator and responder. The firmware is based on the example provided in the chip's Software Development Kit (SDK) ~\cite{qorvofirmware}. The devices perform two SS-TWR rangings per second and use the dynamic STS mode of 802.15.4z HRP for security: with a cryptographic key that was shared out-of-band, they generate an unpredictable, pseudo-random Scrambled Timestamp Sequence (STS) for every new packet. The receiver's code verifies the correctness of the STS by calling the \textit{dwt\_readstsquality} function available in the API. Since the attacker does not know the STS in advance, they cannot advance the packet to reduce the measured distance. Before MD, the only known physical-layer attack possible in this context was GhostPeak, which can cause random distance reductions with a limited success rate~\cite{leu2022ghost}.
We minimally adapted the firmware such that: (i) all parameters can be monitored, (ii) the clock drift of the devices can be trimmed, and (iii) the reply time $T_{reply}$ can be configured.

\begin{figure}[]
    \includegraphics[width=\linewidth]{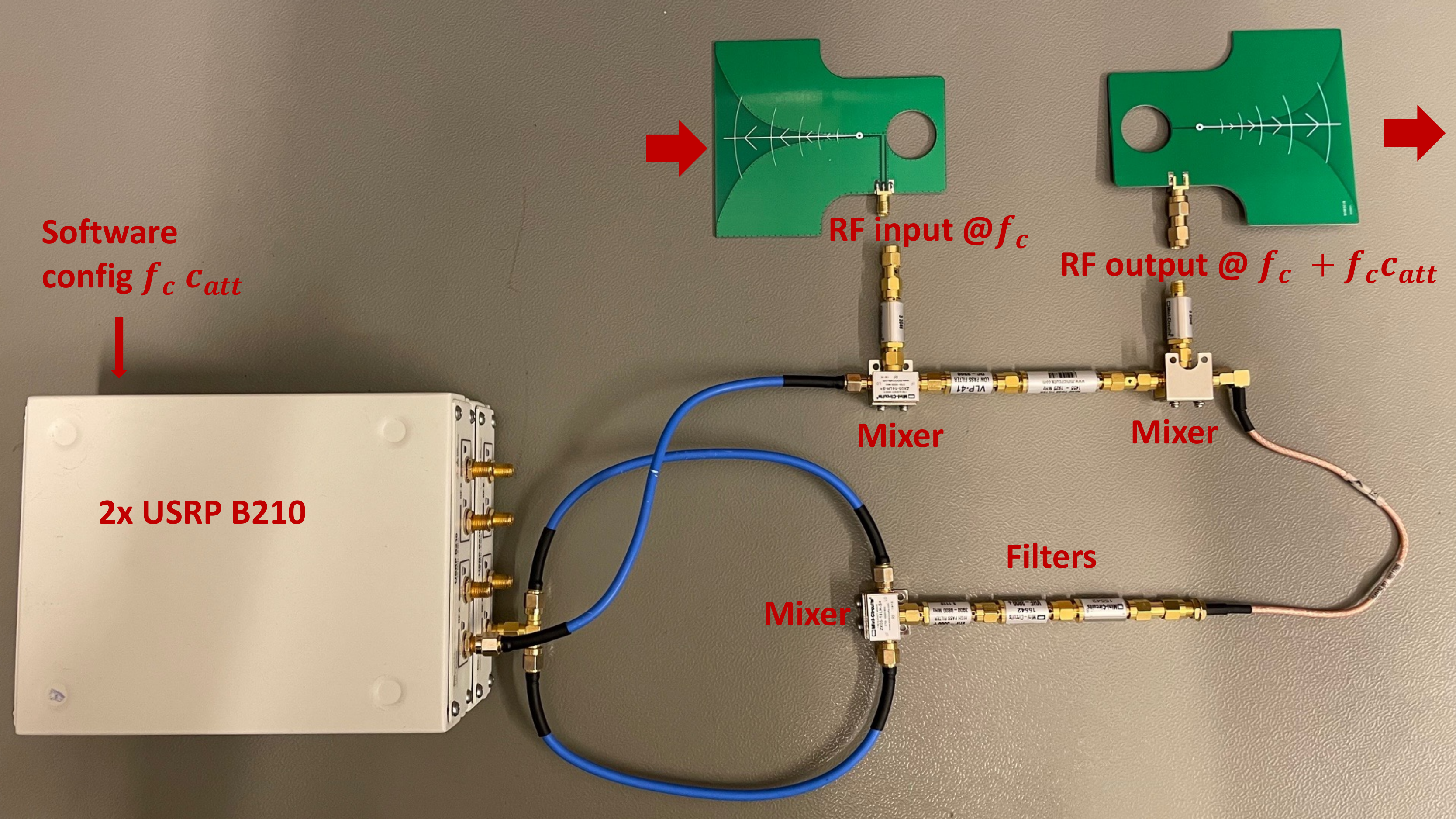}

    \caption{Main components of the attacker. The analog mixing applies the attacker's offset $c_{att}$ to the message from the responder in real-time. The offset can be configured in software since the carriers are generated with USRP B210 SDRs. Antennas can be connected via cables and amplifiers.\label{fig:attacker}}
\end{figure}

\textbf{Attacker:} We implement our attack with a simple analog circuit consisting of off-the-shelf RF components. The clock drift $c_{att}$ can be precisely and flexibly controlled via software. The attacker device is shown in \autoref{fig:attacker}. 
It consists of an antenna to receive the responder's reply at $f_{c}=\SI{6.4896}{\giga\hertz}$ (HRP channel 5), a mixer to convert it down to an intermediate frequency of \SI{1.6}{GHz}, and another mixer to convert it up to the shifted carrier $f_c'=f_c+f_{att}=f_c+f_{c} \cdot c_{att}$.\footnote{Using a single mixer for down-conversion is not possible because the \SI{500}{\mega\hertz} bandwidth of the signal is much larger than the offset.} For example, if an initiator with $c_{init}=0\si{\ppm}$ should measure $c=-10\si{\ppm}$, the attacker applies a frequency shift $f_{att}=-10\cdot$\SI{6.4896}{\kilo\hertz} to bring the carrier down to $f_c'=\SI{6.4895}{\giga\hertz}$. 
The resulting signal is then transmitted over a second antenna in the initiator's direction. We used several amplifiers and filters to improve the quality of the signal. The combined delay caused by the cables and the analog circuitry is below \SI{5}{\ns}. 
The frequencies for down- and upconversion are generated with two USRP B210 Software-Defined Radios (SDRs) and an additional mixer. Apart from the SDRs ($\approx$ USD 1000), which could be replaced by cheaper alternatives, the entire hardware's worth is $\approx$ USD 400.

\textbf{Initial clock drift estimation:}
In \aref{appendix:smooth-takeover-monitor} we present a sniffer that can estimate $c=c_{resp}-c_{init}$ (and/or $c'=c_{att}-c_{init}$) over-the-air, using the messages it receives from the other devices. 
This enables tuning $c_{att}$, for example, for a smooth takevoer in which $c'$ is initially close to $c$.

\textbf{Environment and placement:} We conduct our experiments in an indoor environment. The test devices are stationary during the experiments and up to \SI{10}{\meter} apart. 
They are either in line-of-sight (experiments 1-3) or in different rooms separated by a wall or by glass (experiment 4).

In experiments 1-3, we place the input antenna close to the responder and the output antenna close to the initiator, and use a simple coaxial cable to relay the signals to the attacker. We also relay the first message from the initiator to the responder to facilitate their communication. This setup matches our threat model, which assumes full attacker control over the wireless channel. In experiment 4, we demonstrate that the attack works even if the attacker cannot get close to the responder. 
The sniffer that monitors $c$ and $c'$ is placed in the attacker's vicinity.

The input and output antennas of the attacker are directional and should generally point towards the victims. It is also useful to place them in opposite directions to avoid undesired feedback loops between output and input.

The gain of the attacker can be adjusted from software by controlling the output power of the USRPs. The attacker increases the power until the initiator locks onto their signal.

\textbf{Monitoring and control:} In order to perform systematic and reproducible experiments, we have implemented a control and monitoring system based on the Avatar\textsuperscript{2}~\cite{muench:bar18} framework. Its most important functions are: (i) programming and configuring the victims, (ii) controlling $c_{att}$ programmatically, and (iii) monitoring and storing reception and ranging diagnostics (\textit{e.g.}, measured distance, measured clock drift, STS verification success, and reception errors).

\subsubsection{Experiments and Results}
\begin{figure}[t]
    \includegraphics[width=\linewidth]{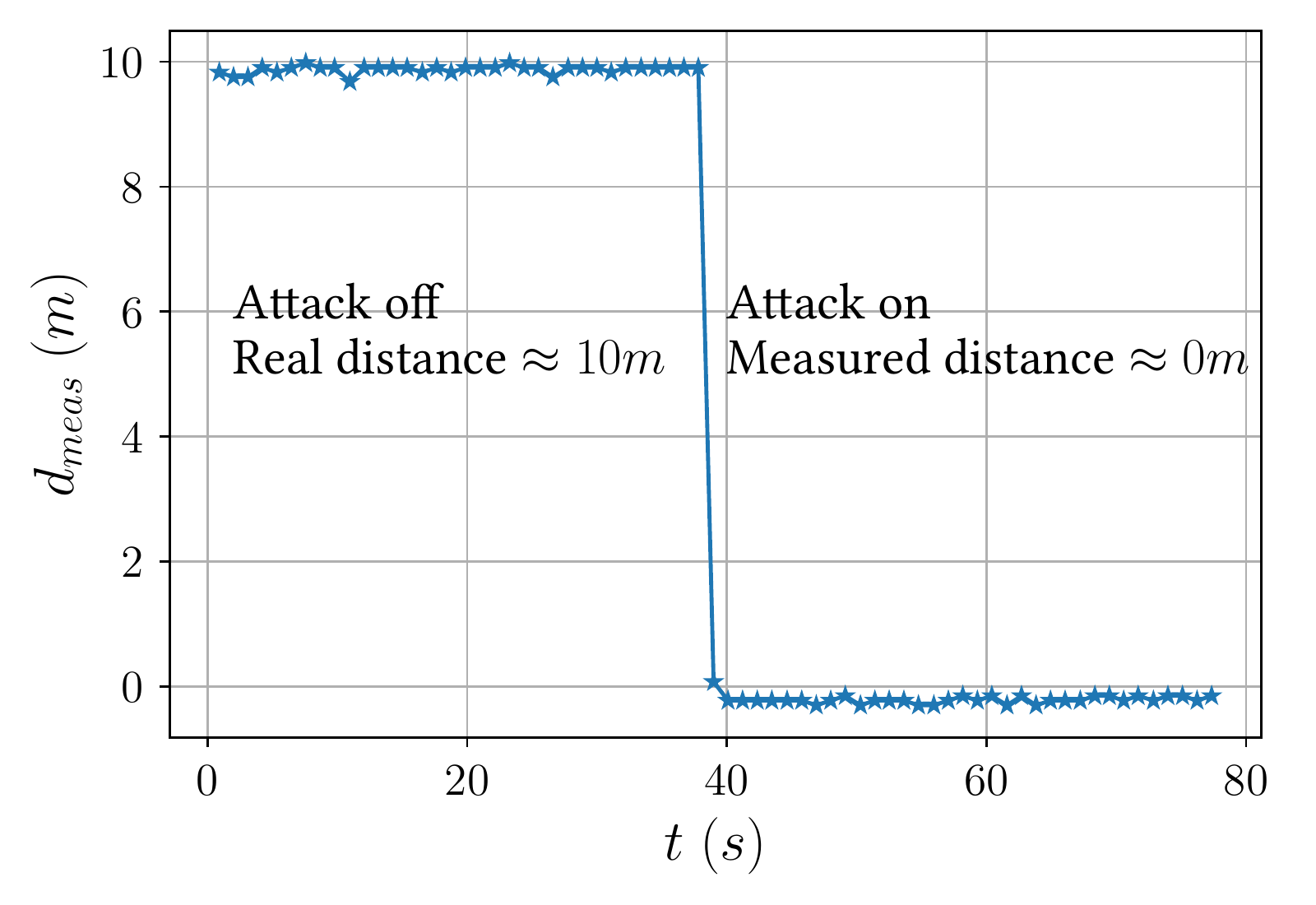}

    \caption{Example of distance reduction from \SI{10}{\meter} to \SI{0}{\meter}.}\label{fig:demonstration}
\end{figure}
\textbf{Experiment 1. Reduction from 10\si{m} to 0\si{m}.} In this experiment, we position the devices in line-of-sight and ten meters apart ($d_{real}=10\si{m}$). Furthermore, we set the reply time to $T_{reply}=5\si{\milli\second}$. 
\autoref{fig:demonstration} shows that once the attack starts at $\approx \SI{40}{\second}$, the measured distance decreases to $\approx \SI{0}{\meter}$ and remains stable at the new value.
As the MD attacker has complete control over the measured distance, it can also obtain the same result with a smooth progressive change (\textit{e.g.}, to simulate a person approaching a car).

\begin{figure}[t]
    \includegraphics[width=\linewidth]{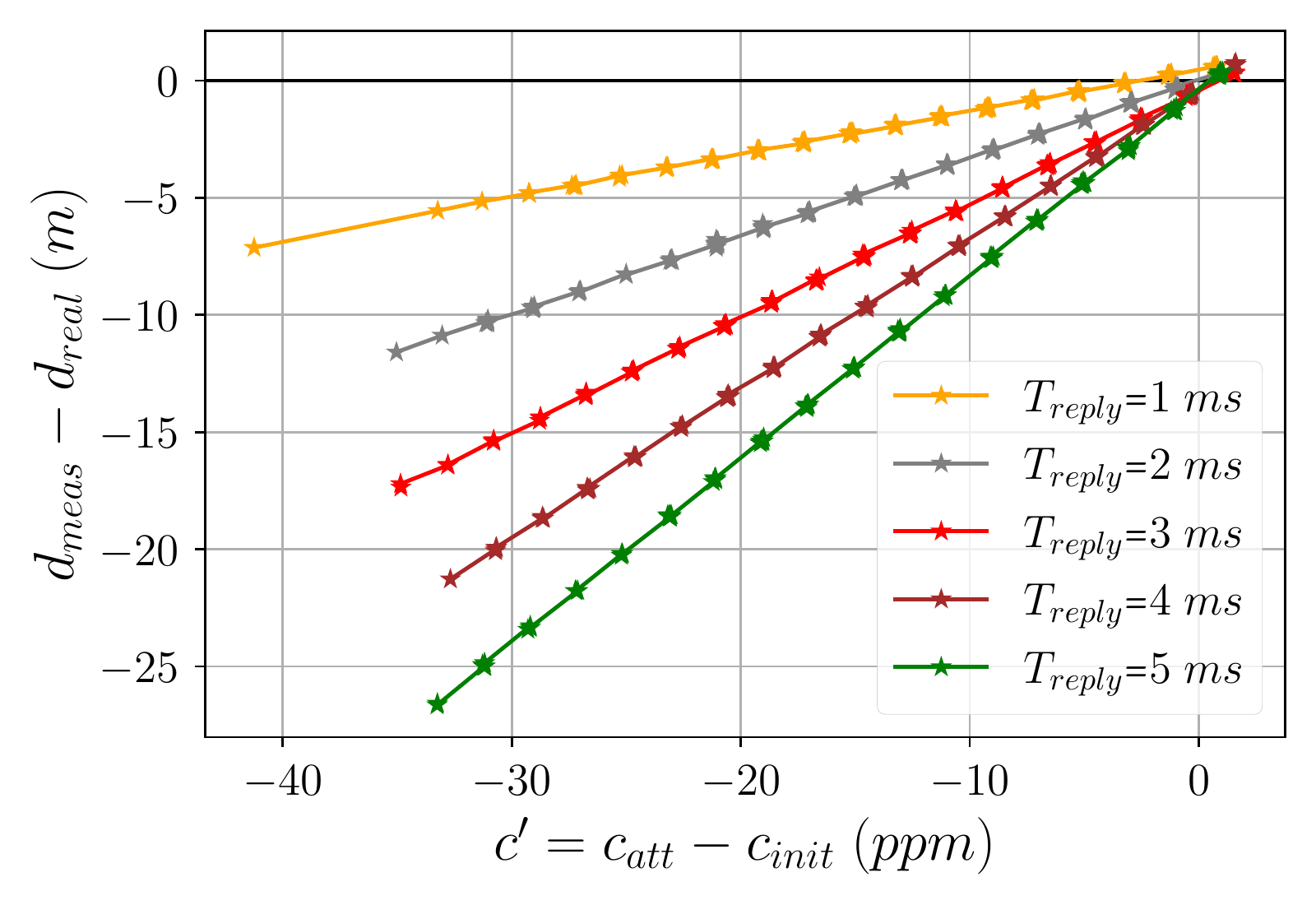}
    \caption{Distance reductions with respect to the attacker's clock drift $c_{att}$ for different reply times. Each star represents a successful measurement. The initator's clock drift $c_{init}$ was constant during the experiment.}\label{fig:delay-systematic}
\end{figure}

\textbf{Experiment 2. Systematic evaluation of $T_{reply}$.}
Neither the DWM3000EVB manual nor 802.15.4z mention any security implications related to $T_{reply}$. While it generally makes sense to keep $T_{reply}$ as short as possible for the sake of measurement accuracy, a well-calibrated initiator (\textit{i.e.}, $c_{init}\approx0\si{ppm}$) can compensate arbitrary errors caused by the responder's clock drift, regardless of $T_{reply}$. Furthermore, use cases such as one-to-many ranging \cite{1tomany} or Real Time Location Service (RTLS) systems with dozens of tags may require longer reply times, since tags reply one after the other.
In this experiment, devices are positioned at a distance of $\approx 2.5\si{m}$. For every value of $T_{reply}\in\{\SI{1}{\milli\second},\dotsc, \SI{5}{\milli\second}\}$ and $c_{att}$ we conduct 10 rangings, and increase $c_{att}$ as long as at least one reception succeeds. The results are shown in \autoref{fig:delay-systematic}.

\textit{Note on proprietary security mechanisms:} The Qorvo DWM3000EVB provides a set of security flags (\textit{e.g.}, "peak growth rate warning" and "ADC count warning"), which are not documented in detail. When the receiver is configured to heed them and discard affected measurements, the MD attack still works. This is expected, since it is unlikely that the warnings are designed to detect our manipulation. However, we observed that the lower bounds of $c'\in [\SI{-42}{\ppm},\SI{-32}{\ppm}]$ (see \autoref{fig:delay-systematic}) were reduced to $[\SI{-25}{\ppm}, \SI{-20}{\ppm}]$. We presume that fine-tuning the attack (\textit{e.g.}, through power adjustments) would mitigate this effect.

\textbf{Experiment 3. Systematic evaluation of $\delta_{c}$.}  We postulate in \autoref{sec:v1_analysis} that the attacker's clock drift advantage $\delta_{c}=c_{att}-c_{resp}$ depends on the clock drifts $c_{init}$ and $c_{resp}$ of the two devices. This is because the initiator should accept an incoming signals if its clock drift estimate $c$ is in the range $[c_{init}-2c_{std},c_{init}+2c_{std}]$. Consequently, both attacker and responder have to be within this range.

To test this assumption, we trim the two transceivers to different clock frequencies using a dedicated device register, and run the MD attack for varying $c_{att}$. We observe that the chip reports reception events for $c_{att}\in[c_{init}-60\si{\ppm}, c_{init}+60\si{\ppm}]$ for any $c_{init}$. This shows that the initiator's clock drift tolerance is even higher than expected, and that the interval is centered around its own clock drift.

However, the initiator only produces a distance measurement if $\delta_{c}\in[-25\si{ppm}, +25\si{ppm}]$ (with small deviations), regardless of $c_{init}$ and $c_{resp}$. Any other $\delta_{c}$ leads to a reception error. Consequently, for the device model that we used, the maximum distance reduction does not depend on the clocks of the ranging devices.

This result can be explained with the sampling frequency offset (SFO, see \autoref{sec:clocks}): the initiator adjusts the sampling rate based on its clock drift estimate $c$. This ensures that, under benign conditions, the signal can be received correctly even for large clock drifts. Under attack however, the initiator adjusts the sampling based on $c_{att}$, whereas the signal was generated using the responder's clock. Therefore, the "correction" by the initiator causes itself an SFO that is proportional to $c_{resp}-c_{att}$. If this difference exceeds a certain threshold, the reception fails.

\textbf{Experiment 4. No physical access to responder.} 
\begin{figure}[t]
    \centering\includegraphics[width=\linewidth]{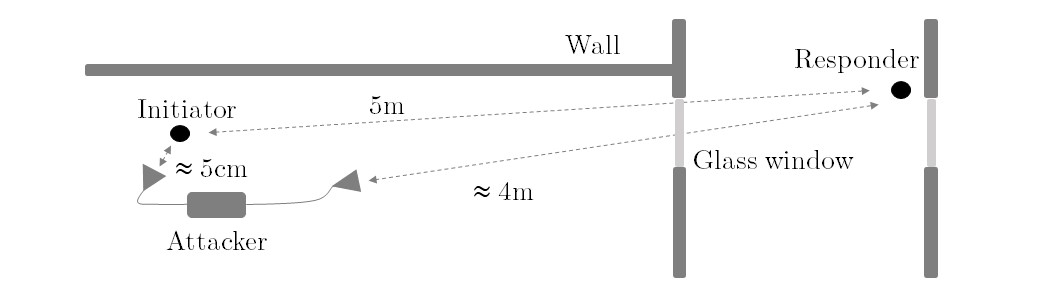}
    \caption{Experiment 4. The responder is placed at $\approx \SI{5}{\meter}$ from the initiator and $\approx\SI{4}{\meter}$ from the attacker, separated by a glass window. The distance estimated by the devices can be reduced to \SI{0}{\meter}.}
    \label{fig:experiment4}
\end{figure}

To demonstrate the feasibility of the MD attack in a more difficult environment, we put the responder into a different room which the attacker cannot access. In two separate runs, the responder is either behind a glass window (see \autoref{fig:experiment4}) or behind a wall. The distance between the attacker's antenna and the responder was approximately $\SI{5}{\meter}$. In this setting, the cable relay used in the previous experiments is not possible, and the received signals have to be amplified to compensate for the additional pass loss. The results show that the distance can still be reduced from $\SI{5}{\meter}$ to $\SI{0}{\meter}$ in this scenario.

\subsection{Countermeasures}
The main problem of SS-TWR is that the initiator has to estimate the responder's clock drift and compensate for it. Without this compensation, it could not provide the accuracy required from UWB ranging. To the best of our knowledge, there is no secure way of verifying the integrity of a CFO under our threat model. Therefore, we argue that MD is hard to prevent, and should instead be mitigated by reducing the reply time and clock drift tolerances. In particular, the following approaches do not promise to resolve the problem:

\textbf{Monitor clock drifts over time:} The devices could track clock drifts to detect attacks. However, a changing clock frequency may have benign causes (\textit{e.g.} device warm-up or temperature changes), leading to a tradeoff between false negatives and false positives.

\textbf{Exchange and compare clock drifts:} During normal operation, the clock drift estimates of the two devices should be almost identical, but with opposite sign. Thus, deviations caused by MD could be detected. However, the attacker could simply manipulate the CFO of both messages accordingly: since the clock drift of the first message is not used in the distance computation, altering it has no consequences, apart from bypassing the countermeasure.

\textbf{Check for presence of second signal:}
If the receiver is able to detect a similar signal at a slightly higher frequency, it could detect the attack. An attacker could bypass this countermeasure by, for example, significantly increasing the power of their own transmission or, if they have sufficient control over the wireless channel, by blocking the genuine signal.

 \section{Attack 2: Stretch-and-Advance (SaA)}\label{sec:v2}

In this section, we present our second attack called Stretch-and-Advance (S\&A). This attack affects the draft of the upcoming 802.15.4ab standard, for which no hardware implementation exists yet. Our evaluation is therefore analytical; for implementation aspects we refer the reader to \aref{appendix:v2-implem}.
We assume the same threat model as for the MD attack, described in \autoref{subsec:threat-model}.

\subsection{Introduction}\label{sec:v2_generalidea}

\begin{figure}[t]
\includegraphics[width=\linewidth]{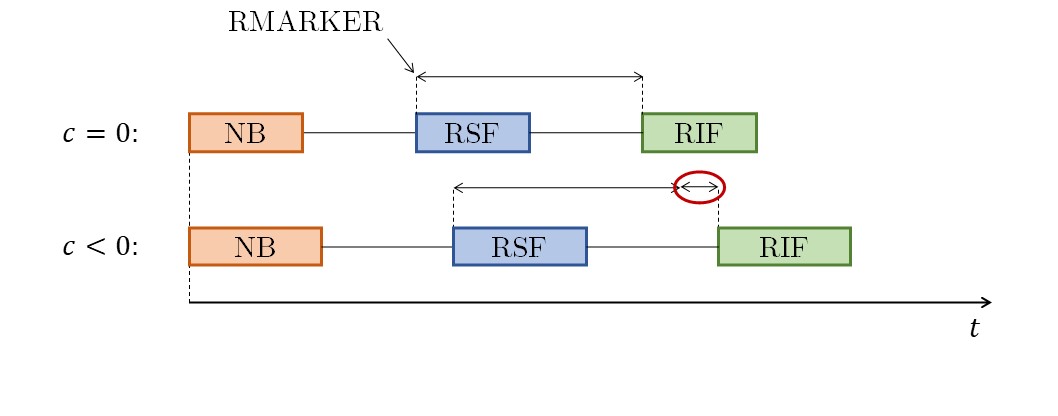}
\caption{Comparison of two ranging messages generated with an ideal clock ($c=0\si{\ppm}$) and a slower clock ($c<0\si{\ppm}$).}
\label{fig:v2_cfo_example}
\end{figure}

A transmitter whose clock frequency is lower than the nominal frequency (\textit{i.e}, with a clock drift $c<0\si{\ppm}$) generates a signal that is "stretched" in time as illustrated in \autoref{fig:v2_cfo_example}. A receiver whose clock operates at the nominal frequency and processes this signal will experience a Sampling Frequency Offset (SFO), since it is in disagreement with the transmitter about the duration of and the delay between UWB pulses. This error grows over time and, without compensation, makes it impossible to acquire and decode long signals. In fact, this phenomenon is responsible for the result of experiment 3 in the \autoref{sec:v1_evaluation} of the MD attack. To avoid this issue, the receiver estimates the clock drift $c$ between the devices by measuring the CFO during the Narrow Band (NB) part of the NBA-MMS-UWB ranging message. Based on the result, it may compensate for the drift either in software or in hardware.

For simplicity, we describe the attack with the basic NBA-MMS-UWB configuration. It contains a single RSF and a single RIF and is shown in \autoref{fig:v2_cfo_example}. The figure illustrates that the delay between the RMARKER and the start of the RIF depends on the transmitter's clock frequency. We will show how this dependency can be exploited to manipulate the ToA of ranging messages, and thus to reduce the distance between the devices. 

\subsection{Attack Description}\label{sec:v2_theattack}
The attacker injects a specially crafted ranging message, which is advanced (\textit{i.e.}, it reaches the receiver before the genuine one) and stretched in time. Because of the stretching, the attacker gains enough time to learn and replay the unpredictable RIF bits from the legitimate transmitter. The resulting message will look to the receiver as if it had been sent by a legitimate transmitter with a slower clock, and its earlier ToA shortens $T_{round}$ (see \autoref{eq:sstwr} and \autoref{eq:dstwr}), which leads to a distance reduction.

\begin{figure}[t]
\centering
    \subfloat[The attacker advances the message by $\delta_{RSF}$, which is the maximum that still allows them to learn and reply all RIF bits in time (the messages are aligned at $t_{replay}$). The advancement depends on the clock drift (the stretch) and the interval between the RMARKER and $t_{replay}$.]{
        \includegraphics[width=\linewidth]{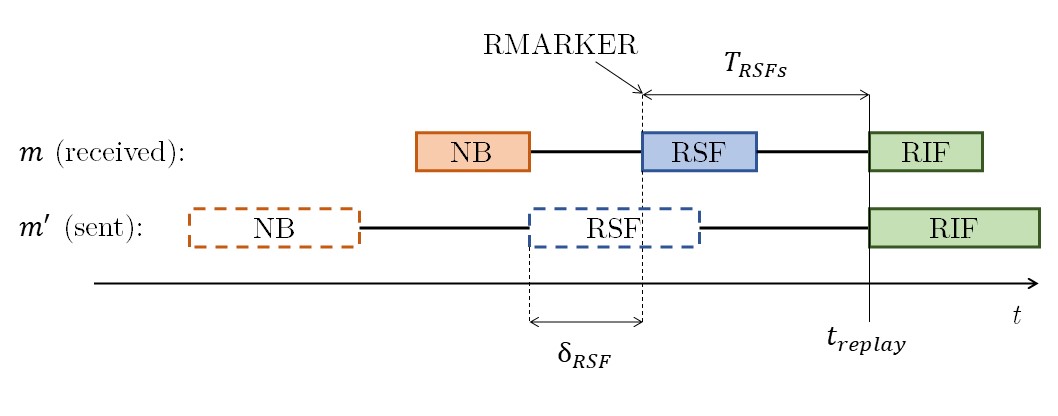}
        \label{fig:v2_attack_a}
    }
    \hfill
    \subfloat[Advancing $m'$ by an additional by $\delta_{RIF}$ is possible, but the leading $n_{att}$ RIF bits (left from $t_{replay}$) become unpredictable.\label{fig:v2_attack_b}]{
        \includegraphics[width=\linewidth]{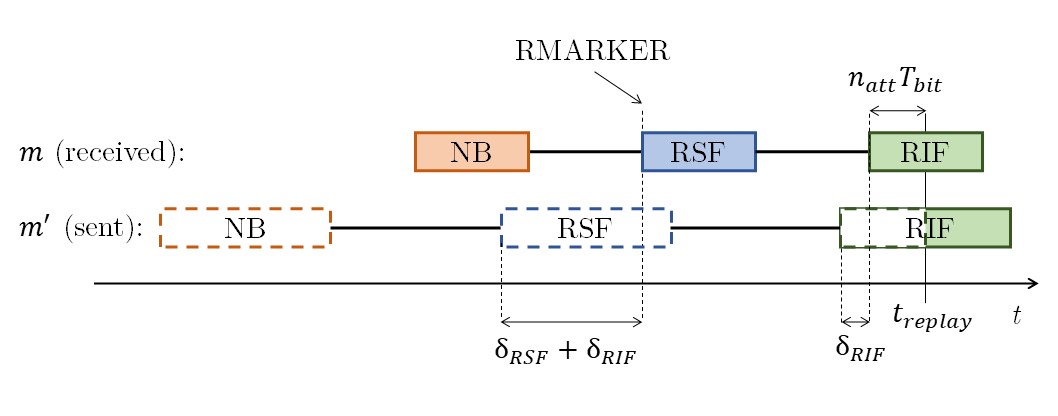}
    }
    \caption{Illustration of the Stretch-and-Advance (S\&A) on a ranging message with a single RIF and a single RSF.\label{fig:v2_attack}}
\end{figure}

\autoref{fig:v2_attack_a} illustrates S\&A from the attacker's perspective. Its upper part shows a ranging message $m$ that would be received from the genuine transmitter, and the lower one the message $m'$ transmitted by the attacker. The attacker constructs $m'$ by first generating the NB and RSF. This is possible because we assume these parts to be deterministic (we discuss non-deterministic NB fragments in \aref{appendix:v2-implem}). The attacker uses a clock with an artificial drift $c_{att}$ in the process, which mimics a slower transmitter and results in the aforementioned stretching. \autoref{fig:v2_attack_a} also shows the advancement or left-shift by $\delta_{RSF}$. This shift is chosen such that the starting times of the RIFs of the two messages ($t_{replay}$) coincide. From $t_{replay}$ on, the attacker can learn and replay the signal and data from the genuine message. By stretching and replaying the entire RIF, the attacker finalizes $m'$.

In the attack description above, we made two implicit assumptions: 

\textbf{Receiver decodes the attacker's message, not the genuine one:} Given that the attacker sends before the genuine transmitter and can choose the transmission power, the receiver will synchronize to the message of the attacker and ignore the legitimate one. In \autoref{sec:v1_evaluation} we have demonstrated that this is possible in practice for the MD attack, even for attacker signals with $\approx 5\si{ns}$ delay.

\textbf{Negligible processing delay:} For simplicity, we assume an attacker capable of receiving and retransmitting RIF pulses instantly, without processing delay. Consequently, the ToF or distance reductions we compute in this section are theoretical upper bounds. In practice, with the analog implementation outlined in \aref{appendix:v2-implem}, we expect negligible delays in the order of nanoseconds, which minimally impact the achievable distance reduction.

We note that the attacker could advance $m'$ by more than $\delta_{RSF}$, as depicted in \autoref{fig:v2_attack_b}. However, this additional shift $\delta_{RIF}$ might come at a cost: it moves $t_{replay}$ into the RIF, which renders the leading RIF bits unpredictable for the attacker. Depending on the value of $\delta_{RIF}$ and the Bit-Error Rate (BER) accepted in the RIF decoding, the success probability of the attack might drop. We leverage this observation in the following security analysis.

\subsection{Analysis}\label{sec:v2_analysis}

The S\&A vulnerability affects SS-TWR and DS-TWR equally. Since both ranging modes average the ToF of all messages (see \autoref{eq:sstwr} and \autoref{eq:dstwr}), we can write the distance reduction  introduced by the S\&A attack as 

\begin{equation}\label{eq:v2_tof_dstwr_raw}
\begin{aligned}
\delta_{ToF} = \frac{a}{N}(\delta_{RSF} + \delta_{RIF})
\end{aligned}
\end{equation}

where $N$ is the number of messages of the ranging scheme and $a$ the number of messages attacked. Consequently, we have $N=2$ for SS-TWR and $N=4$ for DS-TWR\footnote{For the three-message version of DS-TWR, $N=4$ but an attack on the second message counts double, as it affects both ranging rounds.}. In the case of $a\neq1$, this equation assumes that the attacker sets $\delta_{RSF}$ and $\delta_{RIF}$ to be the same across all attacked messages.

In the remainder of the section, we will first quantify $\delta_{RSF}$ and $\delta_{RIF}$. We will then show how $p_{RIF}$, the attacker's probability to pass the RIF integrity check, is coupled to $\delta_{RIF}$.

\subsubsection{$\delta_{RSF}$}
We consider the case shown in \autoref{fig:v2_attack_a}. The advancement of the RMARKER can be written as

\begin{equation}\label{eq:v2_delta_rsf}
\begin{aligned}
\delta_{RSF} &= \delta_{c} \cdot T_{RSFs}
\end{aligned}
\end{equation}
Here, $T_{RSFs}$ is the time duration of all RSFs combined ($N_{RSF}$ fragments at 1\si{ms} intervals), as shown in \autoref{fig:3db_rif}. Like in \autoref{sec:v1}, $\delta_{c}$ denotes the difference in clock drifts between the attacker and the legitimate transmitter. We show in \autoref{sec:v2_std_comp} that an advancement of $\delta_{RSF}$ alone can, when applied to all messages, lead to a distance reduction of over 90\si{m}.

\subsubsection{$\delta_{RIF}$}

If the attacker aims at a larger reduction, they can advance the ranging message by an additional $\delta_{RIF}$ as shown in \autoref{fig:v2_attack_b}. However, this means the leading RIF bits ($\approx 50\%$ in the figure) become unpredictable for the attacker. If the receiver expects all bits of the RIF to be correct, a successful attack would require the attacker to guess these unknown bits. Thus, by increasing $\delta_{RIF}$, the attacker would attempt to further shorten the distance but simultaneously reduce the probability $p_{RIF}$ of passing the RIF verification. We first focus on the advancement $\delta_{RIF}$ and analyze the impact on $p_{RIF}$ subsequently.

We assume that the attacker choses a desired distance reduction, which requires them to apply a given advancement $\delta_{RIF}$ to the ranging message. We denote by $n_{att}$ the number of leading RIF bits made unpredictable by this $\delta_{RIF}$. The longer $\delta_{RIF}$, the higher the number of unpredictable bits the attacker has to accept. This results in

\begin{equation}\label{eq:deltarif}
\begin{aligned}
\delta_{RIF} &= \delta_{c} \cdot \mathcal{T}(n_{att}) 
\end{aligned}
\end{equation}

$\mathcal{T}(n_{att})$ is the time interval corresponding to the $n_{att}$ bits, from the beginning of the first RIF to $t_{replay}$. For our example with a single RIF, we have
\begin{equation}\label{eq:f_simple}
\mathcal{T}(n_{att}) = n_{att}T_{bit} 
\end{equation}

where $T_{bit}$ is the length of a single bit. Generally speaking, $\mathcal{T}(n_{att})$ is a monotonically non-decreasing function of $n_{att}$ that depends on the RIF design and configurations. For ranging messages with several RIFs, $\mathcal{T}$ can be found in \aref{appendix:analysis}.

\subsubsection{Bit errors in RIFs}
For optimal security, a receiver would preferably verify that every single RIF bit is correct. However, multi-path effects, path loss, and transmission power constraints inevitably lead to flipped bits, even in benign environments. Receivers will therefore have to accept RIFs with a non-zero bit error rate. We denote by $BER_{max}$ the highest accepted bit error rate, calculated over all RIFs, below which the receiver accepts RIFs as correct. This assumption is based on the very motivation behind NBA-MMS-UWB: the link budget or the SNR of a single RIF is insufficient to support reliable ranging. Furthermore, the channel might not be coherent over the duration of a long message, especially if the transceivers or objects in their vicinity are in motion.

The number $n_{BER}$ of bit errors accepted by the receiver depends on $BER_{max}$ and $n_{tot}$, the total number of bits in all RIFs:

\begin{equation}
\begin{aligned}
    n_{BER} &= \lfloor BER_{max}\cdot n_{tot}\rfloor
\end{aligned}
\end{equation}

In contrast to standard-compliant transceivers, the attacker can mitigate detrimental channel and noise effects by exploiting their power advantage, favorable positioning, and custom hardware. This enables them to detect and retransmit the \textit{predictable} RIF bits, \textit{i.e.}, those after $t_{replay}$, such that all of them are decoded correctly by the receiver. Therefore, the attacker can, in the best case, make up to $n_{BER}$ guessing errors among the unpredictable bits, and still pass the RIF integrity check. $n_{BER}$ is thus a key parameter that links the reduction $\delta_{RIF}$ to $p_{RIF}$, the probability of passing the check.

\subsubsection{Probability $p_{RIF}$}
Since the attacker can retransmit all predictable bits without errors, they can achieve $p_{RIF}=1.0$ for all $n_{att} \leq n_{BER}$. For $n_{att} > n_{BER}$ however, the attacker is forced to guess at least $n_{att}-n_{BER}$ bits correctly, otherwise the attack fails. The number of bit errors can be modeled with a binomial random variable $X\sim \mathcal{B}(n_{att},0.5)$, and the probability that the attacker makes at most $n_{BER}$ bit errors in $n_{att}$ guesses corresponds to the CDF of $X$ at the position $n_{BER}$, \textit{i.e.},

\begin{equation}
\label{eq:binomprob}
\begin{aligned}
p_{RIF} = P(X \leq n_{BER}) = F_{X}(n_{BER})
\end{aligned}
\end{equation}
$p_{RIF}$ only describes the probability of passing the RIF integrity check of a \textit{single} message for a chosen $n_{att}$. Since an attacker has to pass the integrity check of all $a$ attacked messages (see \autoref{eq:v2_tof_dstwr_raw}), and because the outcomes of these checks are statistically independent events, we can define the overall success probability of the S\&A attack, $p_{succ}$, as

\begin{equation}\label{eq:p_suc}
     p_{succ}=\prod_{i=1}^{a}{p_{RIF}}_i
 \end{equation}

\subsection{Impact on UWB standards}\label{sec:v2_std_comp}

\begin{figure*}[t]
\includegraphics[width=\linewidth]{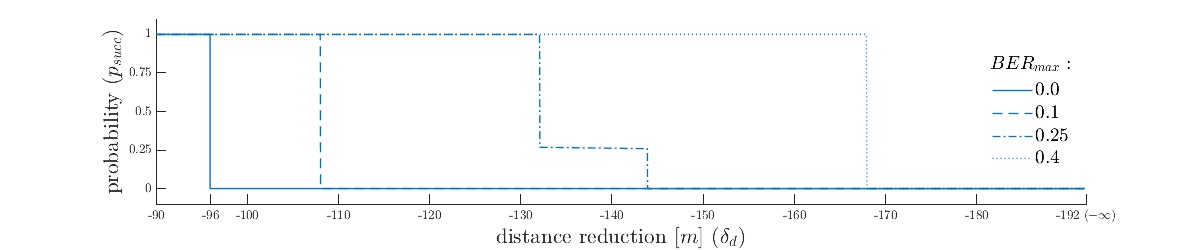}
\caption{The success probability $p_{succ}$ depends on the distance reduction $\delta_{d}$ (\textit{i.e.}, $\delta_{ToF}$ times the speed of light). Every curve corresponds to the results for a specific $BER_{max}$. The attacker can always achieve $\delta_{d}=-95.93\si{m}$, even if the receiver tolerates no bit errors. The discontinuity for $BER_{max}=0.25$ is related to the pauses between RIFs, and enables the attacker to increase the distance reduction, to some extent, without sacrificing success probability (see the definition of $\mathcal{T}$ in \aref{appendix:analysis}). A distance reduction of about -192\si{m} requires an advancement $\delta_{RIF}$ that renders all RIF bits unpredictable.}
\label{fig:v2_analysis_4ab}
\end{figure*}

The maximum distance reduction depends on several parameters. To estimate the effect of S\&A on 802.15.4ab, we use the configuration introduced in \autoref{sec:mms_uwb}. For 802.15.4z we use the default configuration of the HRP mode.

\subsubsection{802.15.4ab}\label{sec:saa-newstd}
We have conducted the evaluation for ranging messages with 8 RSFs and 8 RIFs, each at $1\si{ms}$ intervals. Every RIF comprises $n_{RIF} = 512$ bits, which adds up to $n_{tot}=4096$ bits. Within a RIF, bits are sent at $T_{bit}=32\si{ns}$ intervals. Like in \autoref{sec:v1_analysis}, we assume $\delta_{c}=-40\si{ppm}$ to compute reduction limits. We assume that the devices use SS-TWR, and that the attacker advances both messages equally to optimize $p_{succ}$.

\textbf{Results:} Our evaluation shows that the distance reductions possible with the S\&A attack are large enough to defeat any proximity-based access control system, irrespective of the tolerated bit error rate. 

\autoref{fig:v2_analysis_4ab} shows the success probability of the S\&A attack in dependence of the distance reduction. The attack is deterministic ($p_{succ} = 1.0)$ for reductions of at least $\delta_{d}=-93.95\si{m}$, which can be achieved through advancing both messages by $\delta_{RSF}$ only (see \autoref{eq:v2_delta_rsf}). 
Longer reductions require $\delta_{RIF}< 0$ and result in unpredictable RIF bits. Nevertheless, the distance between two transceivers tolerating $BER_{max}=0.1$ can be reduced by $\delta_{d}=-107\si{m}$ while still keeping $p\approx 1.0$. For $BER_{max}=0.25$, reductions of $\delta_{d}=-132\si{m}$ ($p_{succ}= 1.0$) or $\delta_{d}=-143\si{m}$ ($p_{succ}\approx 0.5$) are possible. For simplicity, we rounded all distance reductions to meters.

\subsubsection{802.15.4z}
The current 802.15.4 standard is practically unaffected by the S\&A attack: generally speaking, 4z messages are too short for exploitable attack margins.

Out of the four message formats defined for HRP (see~\cite{9179124}, Figure 15-2a), two put the RMARKER only microseconds before the start of the STS, which is the HRP equivalent of the RIF. Hence, the stretchable timeframe in \autoref{eq:v2_delta_rsf} is too short and only results in a negligible $\delta_{RSF}<1\si{ns}$. Another message format puts an encrypted payload field between the RMARKER and the STS, which is unpredictable and cannot be advanced, effectively leading to the same result. The last format does not contain any STS field at all, has no security claims, and can be attacked with more trivial means.

For $\delta_{RIF}$, the situation is similar: the STS is comparably short (\textit{e.g.}, $\approx 131 \si{\micro\second}$ for $n_{tot}=8192$ pulses). Even in the extreme case, where the attacker chooses $t_{replay}$ such that $n_{att} \approx n_{tot}$, $\delta_{RIF}$ would be negligible:
$\delta_{RIF} = 131 \si{\micro\second} \cdot -\SI{40}{ppm} = -\SI{5.24}{ns}$.

More exotic, optional configurations exist (\textit{i.e.}, multiple and longer STS segments with additional RMARKERS) but the standard does not detail how they should be processed. To the best of our knowledge, we therefore conclude that distance reductions with S\&A are negligible in 4z HRP.

\subsection{Countermeasure}\label{sec:v2_countermeasures}
In this section, we propose a countermeasure that allows to bound the success probability of S\&A distance reductions to an arbitrarily low value. 

\subsubsection{Working Principle}

The S\&A attack can be mitigated if the two devices check if their clock drift estimates are consistent. Under benign conditions, a relative clock drift $c$ measured by the responder causes the initiator to report approximately $-c$. In contrast, the S\&A attack described before would cause a conspicuous deviation. The countermeasure we propose prevents S\&A and cannot be bypassed if receivers set $BER_{max}\approx0.2$. 

We assume that both devices estimate their clock drift based on the CFO of the NB fragment before the UWB exchange begins, and use this estimate for the duration of the entire reception. We do not detail how the clock drift estimates have to be exchanged, but we require their integrity to be cryptographically protected. Both conditions should be easy to fulfill: the devices need the means to exchange timing information for DS-TWR in any case, and the generation of RIF sequences already requires a shared secret.

\subsubsection{Security Analysis}
For the sake of simplicity, we analyze the countermeasure for SS-TWR and the basic NBA-MMS-UWB configuration with one RIF. 
We transfer the result to DS-TWR at the end of the section, and to the case of multiple RIFs in \aref{appendix:analysis}.

The following analysis is based on the publicly available draft documents and proposals of 802.15.4ab at the time of writing. Since they may differ from the final specification, the proposal should go through the full standardization process.

In SS-TWR, initiator and responder send one message each (\textit{i.e.}, $m_1$ and $m_2$) and produce a clock drift estimate on the received message. In order to achieve a distance reduction, the combined ToF change of both messages must be negative:
\begin{equation}
\begin{aligned}
  \delta_{ToF} &= \frac{1}{2}(\delta_{ToF1} + \delta_{ToF2})\footnotemark\\
\end{aligned}
\footnotetext{The division by 2 is due to the averaging in \autoref{eq:sstwr}.}
\end{equation}

Without loss of generality, we assume that the attacker successfully conducts the S\&A attack on $m_{1}$. In the process, they change the clock drift estimate of the responder by $\delta_{c}<0$ and achieve a ToF reduction of

\begin{equation}\label{eq:cm_deltatof1}
\begin{aligned}
   \delta_{ToF1} &=\delta_{c}\big(T_{RSFs}  + \mathcal{T}(n_{att1})\big)
\end{aligned}
\end{equation}

\begin{figure}
    \includegraphics[width=\linewidth]{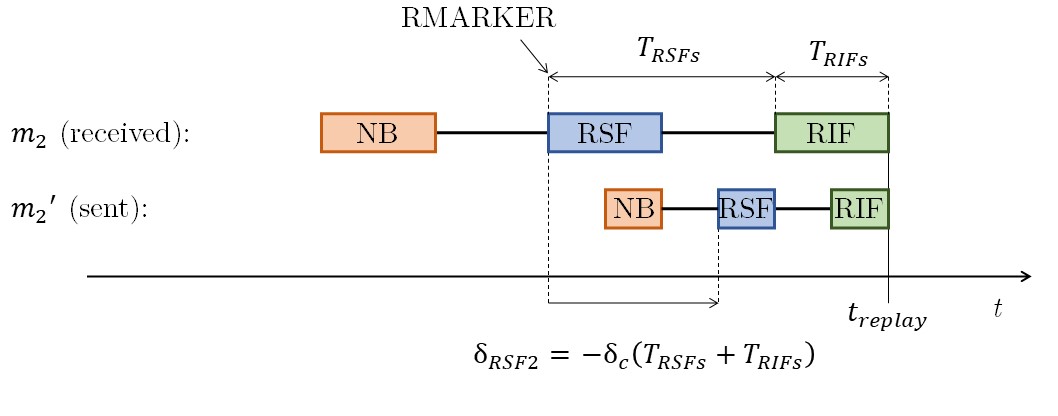}
    \caption{To replay all RIF bits, the messages have to be aligned at the end of the RIF, which shifts the RMARKER by $\delta_{RSF2}$ to the right and increases the ToF compared to $m_{2}$.}
    \label{fig:v2_countermeasure_m2}
\end{figure}

In order to bypass the countermeasure, the attacker also has to manipulate the initiator's clock drift estimate  by $-\delta_{c}>0$. In this case, the estimates are consistent and the transceivers accept the ranging result. However, this causes the initiator to expect a compressed message, because it concludes that the responder's oscillator is too fast. Because of the resulting sampling frequency offset (SFO, see \autoref{sec:clocks}), the reception of $m_{2}$ inevitably fails. This already happens in 802.15.4z for much shorter messages  (see experiment 3 in \autoref{sec:v1_evaluation}). To prevent this, the attacker has to forge a compressed $m_{2}'$. This means they have to send the RIF pulses at a higher rate than they can be received from the genuine transmitter. To replay all RIF bits correctly, the attacker has to delay the transmission of $m_{2}'$ long enough to learn the last RIF pulse just in time for retransmission. This is depicted in \autoref{fig:v2_countermeasure_m2}, which illustrates how compressing and delaying $m_{2}'$ increases its ToF by $\delta_{RSF2} = -\delta_{c}(T_{RSFs} + T_{RIFs})$.

If the attacker is willing to compromise on the RIFs correctness, they can advance $m_{2}'$ and mitigate the enlargement. The advancement can be written as $\delta_{RIF2}=-\delta_{c}\mathcal{T}(n_{att2})$, where $n_{att2}$ is the number of \textit{trailing} RIF bits that become unpredictable. This results in

\begin{equation}\label{eq:cm_delta_tof2}
\begin{aligned}
\delta_{ToF2} &=-\delta_{c}\big(T_{RSFs} + T_{RIFs} - \mathcal{T}(n_{att2})\big)
\end{aligned}
\end{equation}

Since we consider $N_{RIF}=1$, we can use $T_{RIFs} = n_{tot}T_{bit}$ and $\mathcal{T}$ as defined in \autoref{eq:f_simple}, which finally results in

\begin{equation}\label{eq:tof_cm}
    \begin{aligned}
        \delta_{ToF} = \frac{\delta_{c}T_{bit}}{2}(n_{att1}+n_{att2}-n_{tot})
    \end{aligned}
\end{equation}

\begin{figure}[t]
    \includegraphics[width=\linewidth]{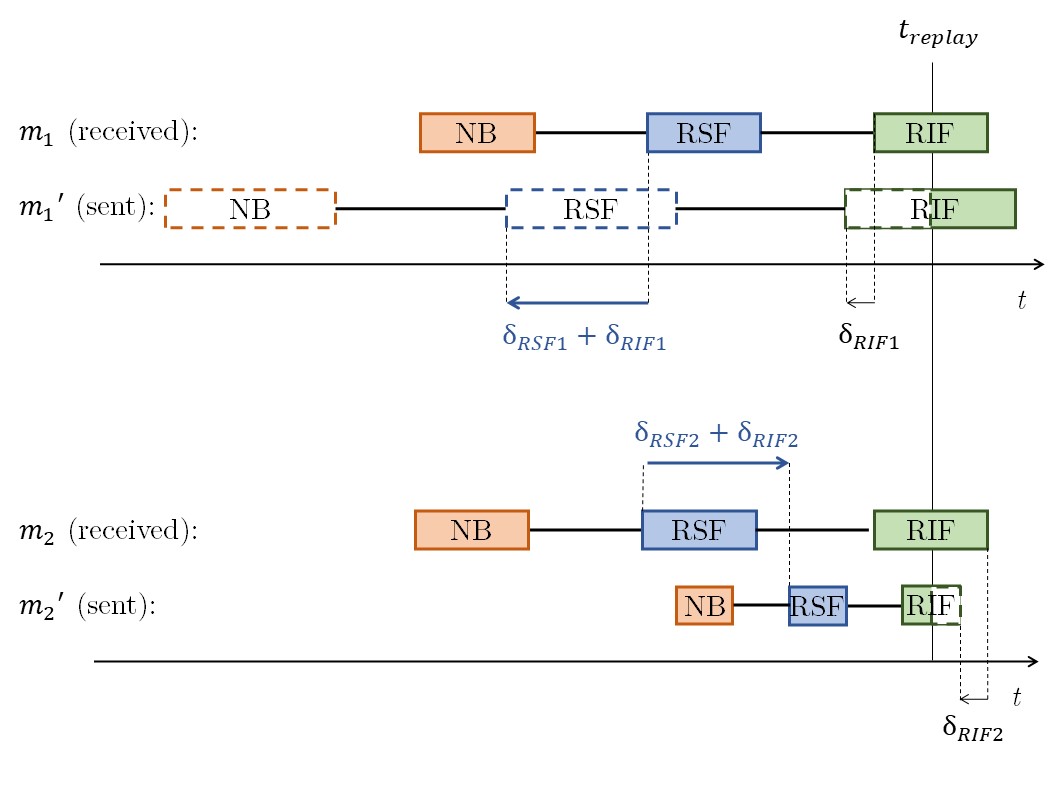}
    \caption{Despite the countermeasure, the attacker attempts to reduce the distance with the S\&A attack. They stretch $m_{1}'$ and compress $m_{2}'$, and advance both of them as shown. As a result, half of the RIF bits become unpredictable. However, the advancement of $m_{1}'$ and the delay of $m_{2}'$ (blue arrows) have the same value, \textit{i.e}, the attack does not change the computed ToF/distance. Further advancements would result in a distance reduction, and must be prevented by the RIF integrity checks.}
    \label{fig:v2_countermeasure_2}
\end{figure}

Therefore, the attacker can reduce the distance if they accept $n_{att1}+n_{att2}>n_{tot}$ unpredictable RIF bits. \autoref{fig:v2_countermeasure_2} illustrates  how $n_{att1}=n_{att2}=n_{tot}/2$ pushes $t_{replay}$ into the middle of the RIFs of both messages, which compensates for the delay $\delta_{RSF2}$. 

Since any further shifts would result in a distance reduction,  we want to ensure that setting $n_{att1}+n_{att2}\geq n_{tot}$ makes it hard for the attacker to pass the RIF integrity checks\footnote{for $N_{RIF}>1$ the equality already leads to a reduction, but any $n_{att1}+n_{att2}<n_{tot}$ does not. We refer readers to \aref{appendix:analysis} for more information.}. Specifically, we want to ensure that choosing $n_{att1}=n_{att2}=n_{tot}/2$ is infeasible, since this case maximizes the attacker's success probability $p_{succ}$ for any sum  $n_{att1}+n_{att2}=n_{tot}$ (see \autoref{eq:binomprob} and \autoref{eq:p_suc}). 

\begin{figure}[t]
    \includegraphics[width=\linewidth]{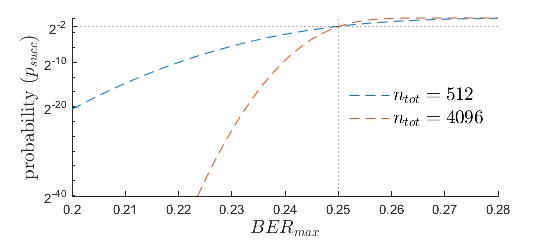}
    \caption{Probability $p_{succ}$ of passing the RIF integrity checks in both SS-TWR messages for $n_{att1} = n_{att2} = n_{tot}/2$. The probability depends on the overall number of bits all RIFs combined. $n_{tot}=512$ corresponds to one RIF, $n_{tot}=4096$ to 8 RIFs.}
    \label{fig:v2_ber_epsilon}
\end{figure}

\textbf{Results:}
\autoref{fig:v2_ber_epsilon} shows how a receiver can reduce the success probability by reducing the accepted bit-error rate. We see that the success probability is sensitive to small changes in $BER_{max}$. For $n_{tot}=4096$, $BER_{max}=0.22 \Rightarrow p_{succ}\leq2^{-40}$, and $BER_{max}=0.25 \Rightarrow p_{succ}=2^{-2}$. The reason for the latter is that an attacker with $n_{att1}=n_{att2}=n_{tot}/2$ has a $p_{RIF}=0.5$ of passing the RIF check for each message, resulting in $p_{succ}=0.25$ for two messages.

We have shown that the clock drift consistency check protects SS-TWR against S\&A. Since DS-TWR comprises two SS-TWR rounds, it is equally secure if the clock drifts of all involved messages are compared. In fact, it is even harder for an attacker reduce the distance in DS-TWR because they have to pass the RIF integrity check in four messages instead of two.

\subsubsection{Further countermeasures and future work}
Executing S\&A with $\delta_{RIF}<0$ inevitably results in an uneven distribution of bit errors between the predictable and unpredictable parts of the RIF (\textit{i.e.} up to $BER=0.0$ vs $BER\approx0.5$ for an ideal attacker, respectively). Receivers could analyze this distribution as an additional hardening measure. However, this could result in a classification problem of distinguishing between malicious behavior and bad channels, and would require a thorough analysis.
Another, simple countermeasure worth evaluating is to move the RMARKER's position into the middle of the RIFs. This would categorically prevent $\delta_{RSF}<0$ and, thus, limit the maximum distance reduction significantly. However, distance reductions due to $\delta_{RIF}<0$ would still be possible, and likely be subject to similar $BER_{max}$ requirements like the countermeasure we have proposed.


\section{Related Work}

\textbf{Distance bounding:} 
Distance bounding protocols~\cite{DBLP:conf/eurocrypt/BrandsC93, DBLP:conf/wisec/TippenhauerLKC15,DBLP:conf/wpnc/KuhnLT10} establish an upper bound of the distance between a prover and a verifier under various threat models (\textit{e.g.}, external attacker, honest/dishonest prover).
Practical realizations of distance bounding protocols based on wireless ranging have been proposed (\textit{e.g.}, based on UWB~\cite{DBLP:conf/uss/RasmussenC10} or NFC~\cite{DBLP:conf/securecomm/HanckeK05}).
In practice, secure ranging in the UWB standard~\cite{9144691,9179124} assumes honest initiators and responders that protect ranging integrity using pseudo-random sequences in the ranging frames.
A survey on distance bounding protocols and attacks can be found here~\cite{DBLP:journals/csur/AvoineBBCHKKLMM19} .

Our work is orthogonal to the design of distance bounding protocols because it focuses on the manipulation of time references at the physical layer.

\textbf{Physical-layer attacks:}
Guaranteeing the integrity of distance measurements against physical-layer attacks is not trivial. 
Systems based on signal strength or phase are insecure~\cite{DBLP:conf/ndss/FrancillonDC11,DBLP:conf/ches/OlafsdottirRC17}. Chirp-based systems are vulnerable to Early-Detect/Late-Commit attacks~\cite{DBLP:conf/wisec/RanganathanDFC12}, which also affect Ultra Wide Band Impulse Radio (802.15.4a)~\cite{DBLP:journals/twc/PoturalskiFPHB11}. In Cicada attacks~\cite{DBLP:journals/twc/PoturalskiFPHB12,5616900}, the receiver mistakes the attacker's random pulses for a weak legitimate signal. The HRP mode of 802.15.4z is vulnerable to similar and improved attacks~\cite{DBLP:conf/wisec/SinghRZLC21,leu2022ghost}.

\textbf{Time-of-Arrival Integrity:}
Leu et al. introduced Message Time-of-Arrival Codes (MTACs)~\cite{DBLP:conf/sp/LeuSRPC20}, a framework to formalize the security requirements of ToA measurements, taking into account pyshical layer attacks such as Early-Detect/Late-Commit (ED/LC). However, they do not consider the possibility of an attacker manipulating the time references of involved devices.

\textbf{Clock imperfections:}
From a functional point of view, clock non-idealities in wireless ranging systems are well understood, and compensation techniques exist~\cite{DBLP:conf/wpnc/DotlicCM18,DBLP:conf/wpnc/NeirynckLM16}.
The formulas described in 802.15.4z~\cite{9144691,9179124} minimize the effect of clock drift for both SS-TWR and DS-TWR. The 802.15.4ab standard~\cite{abdocs} will require highly accurate CFO estimates to support multi-millisecond frames.

Clulow et al.~\cite{DBLP:conf/esas/ClulowHKM06} propose an attack on ToF distance bounding in which a malicious prover leverages packet-level latencies to reply earlier to the challenge sent by the verifier. The authors also mention overclocking the prover to further shorten its reply time, relying on the tolerance of the receiver for clock drifts. In this attack, the malicious prover actually runs faster to shorten its reply time. In contrast, in MD, an external attacker convinces the initiator that the responder was slower, so that the initiator compensates for a larger reply time and computes a shorter distance. With clock drift compensation in place, overclocking the responder would have no effect. 

In the context of OFDM-based ranging, Singh et al.~\cite{DBLP:conf/ndss/SinghRRC22} propose a "carrier frequency offset attack" for distance enlargement. The attacker overshadows the reference signal with a wrong carrier frequency offset, preventing decoding of legitimate data, and replays the legitimate signal with a delay, causing a distance enlargement.

Non-ideal clocks are a problem also in sensor networks where each device has a defined time slot to transmit. In this context, algorithms for secure time synchronization have been proposed in literature~\cite{DBLP:journals/tissec/GaneriwalPCS08}.

\section{Conclusion}
Because of their non-ideal clocks, different electronic devices have a different notion of time and frequencies. Distance measurement based on the time of flight works under the implicit assumption that initiator and responder can deal with such differences through various forms of compensation.
While these aspects are well known from a functional point of view, they have been largely overlooked in the security analysis of ranging.
This is a topical issue in the definition of the upcoming 802.15.4ab standard, whose long frame duration exacerbates the problems of non-ideal clocks, but it is already relevant to the current 802.15.4z standard. 

In this paper we provided an extensive analysis of the security impact of clock non-idealities and their compensation. We have presented two over-the-air attacks, Mix-Down and Stretch-and-Advance, which compromise the integrity of UWB measurements and lead to substantial distance reductions. This is a grave concern in security-critical applications, such as proximity-based access control. Our proposed countermeasures can defend against S\&A, while MD does not have a straightforward solution.

Our insights prove that the implications of physical layer aspects on ranging integrity are still not completely understood, and should be analyzed in the design of new standards and protocols.

\section{Acknowledgements}
This research has received funding from the
Swiss National Science Foundation under NCCR Automation, grant agreement 51NF40\_180545.





\bibliographystyle{plain}
\bibliography{references}

\appendix
\section{Appendix}
\subsection{Over-the-Air Estimation of Clock Drifts}
\label{appendix:smooth-takeover-monitor}

We show that an attacker can estimate clock drifts over the air, using only the messages it receives from other devices, and leverage this information to fine-tune its own clock drift.
In particular, this would enable smooth takeover attacks in which $\delta_{c}=c'-c=c_{att}-c_{resp}$ is initially set to $\approx \SI{0}{\ppm}$ and then gradually decreased.

The attacker estimates the natural drift $c=c_{resp}-c_{init}$ between initiator and responder by simply sniffing their messages with a receiver, and computing
\begin{equation}
    c = c_{resp}-c_{init} = (c_{sniffer}-c_{init}) - (c_{sniffer}-c_{resp}) 
\end{equation}
where $c_{sniffer}-c_{init}$ ($c_{sniffer}-c_{resp}$) is the drift estimated by the sniffer when receiving a message from the initiator (responder). 
The location of the sniffer with respect to the other two devices does not affect the computation as the arrival time is not used.
An additional rolling median filter over a sliding window can be applied to filter noise and outliers.

Assuming that the initiator is well calibrated and has a negligible drift, the attacker can choose $c_{att}=c\approx c_{resp}$. 
If this is not the case, the attacker could use the same method to estimate $c'=c_{att}-c_{resp}$, by placing a second sniffer directly at the output of its mixing stages.
The attacker could then start with a low power, adjust the mixing frequency until $c'=c$ (i.e., $c_{att}=c_{resp}$) and then start increase the power to takeover. 
The advantage of this method is that it is a self calibration of the attacker to the responder over-the-air.

To demonstrate the ability of the attacker to estimate clock drifts over-the-air with good accuracy, we have implemented a sniffer with a QorvoDWM3000 transceiver, and we have evaluated it in a scenario without physical (the same as experiment 4 described in \autoref{sec:v1_evaluation}).

\begin{figure}[t]
    \includegraphics[width=\linewidth]{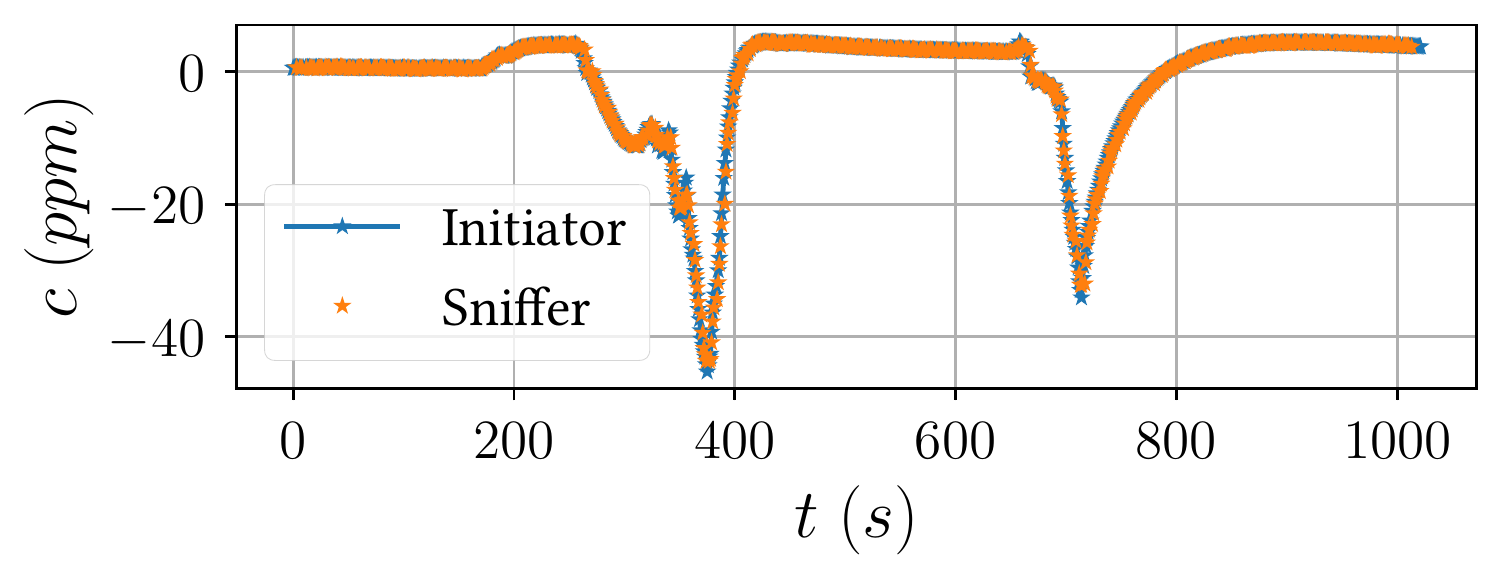}
    \caption{The MD attacker uses a sniffer to estimate the natural clock drift between intitiator and responder. The estimate of the sniffer (orange) follows closely the one of the initiator (blue) shown as ground truth.}\label{fig:cfo_monitor}
\end{figure}

\autoref{fig:cfo_monitor} shows how the value estimated by the attacker over the air closely follows the value actually estimated by the initiator. The larger variations at around \SI{200}{\second} - \SI{600}{\second} and \SI{800}{\second} - \SI{400}{\second}  were caused by a sudden increase in the temperature of the initiator that we induced with an off-the-shelf hair-dryer. The sniffer was able to follow the variations.

\subsection{Implementation of S\&A}
\label{appendix:v2-implem}

\begin{figure}[t]
    \includegraphics[width=\linewidth]{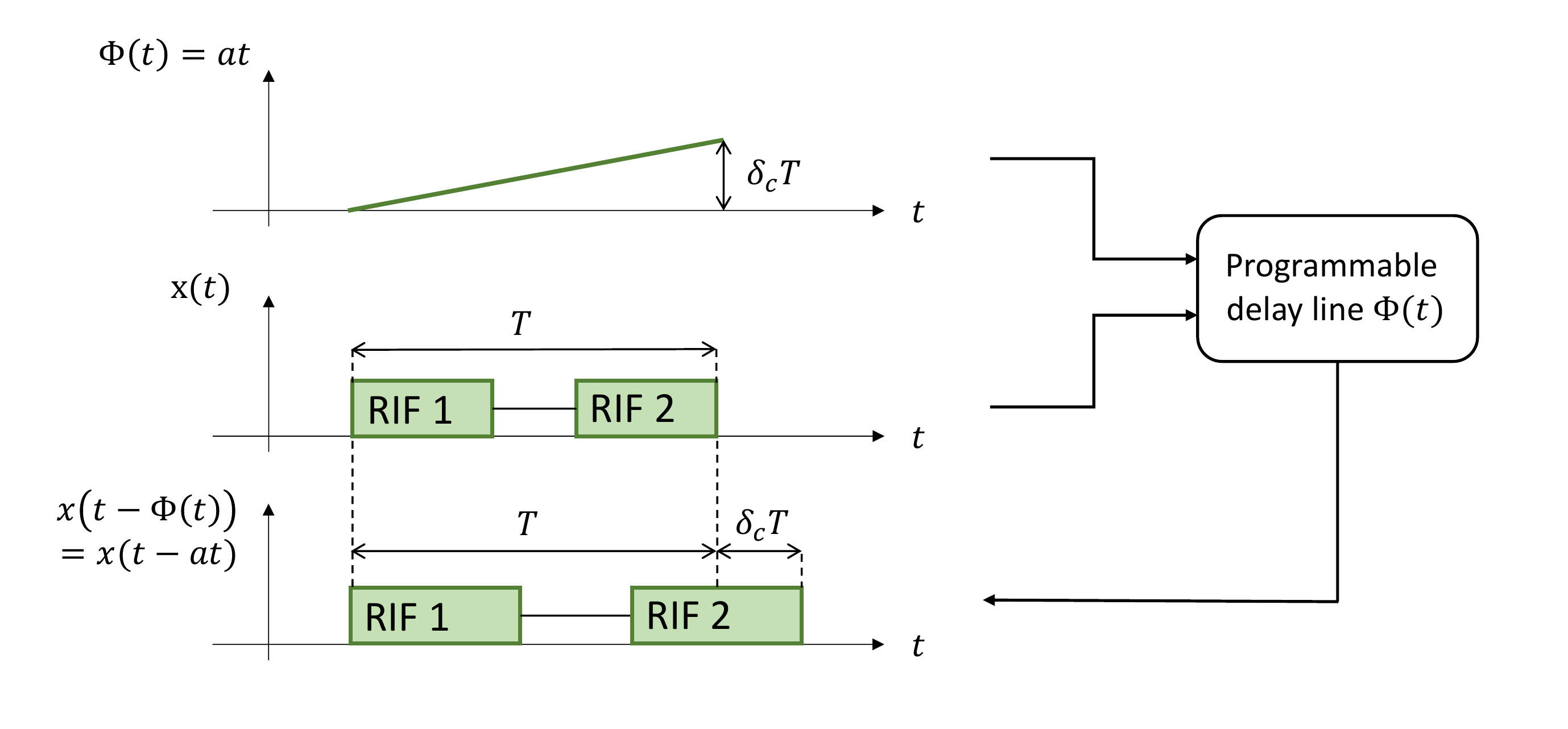}
    \caption{Proposed architecture to replay the unpredictable RIF bits with low latency in the S\&A attack: the attacker stretches a signal in real-time by applying a linearly increasing delay with a programmable delay line (e.g., based on~\cite{PhysRevX.9.031015,9354215}).}\label{fig:programmable-delay-line}
\end{figure}

In contrast to MD, S\&A requires signals to be stretched in time. This has to be done differently for predictable and unpredictable fragments, but the two approaches can be combined into one attacker device.

\textbf{RSFs and NB:} The RSF and NB fragments can be generated and stretched using an off-the-shelf UWB transceiver. The stretching can be achieved by trimming the transceiver's clock in hardware or in the firmware. \footnote{For the current standard, the Qorvo DMW3000EVB offers clock trimming from software, and we expect future devices to have similar features.}

\textbf{RIFs:} Since they are unpredictable, the attacker has to receive the RIF pulses and retransmit them with increasing delay. This could be implemented with programmable delay lines, based on UWB switched-capacitor delay elements~\cite{PhysRevX.9.031015,9354215}. \autoref{fig:programmable-delay-line} shows how a delay line works: an analog circuit applies the delay $\Phi(t)$ to an input signal. Although $\Phi(t)$ is increased in small digital steps, the resolution is high enough to assume it to be linear. The authors have built a working prototype that delays 500\si{MHz} UWB signals several dozens of nanoseconds, which is the order of magnitude required for the RIFs.

\textbf{Unpredictable NB:} If the NB fragment contains unpredictable data (\textit{e.g.}, cryptographic nonces), the previous attack strategy has to be revised. Here, the PHY layer properties of the NB fragment can be helpful: according to the standard draft, the NB channels only have a bandwidth of 2.5\si{MHz}, which results in a pulse length of $\approx1\si{\micro\second}$. The attacker could exploit this twofold:

First, note that we exaggerated the effects of stretching and advancing in our figures. They would be imperceptible if shown to scale: stretching a 17\si{ms}-long ranging message with $\delta_{c}=-40\si{ppm}$ only increases its length by 0.000'68\si{ms}. This is the maximum by which the NB fragment has to be advanced throughout all configurations. We expect that, after receiving the NB fragment, the receiver accepts the first RSF to arrive a few hundreds of nanoseconds earlier or later, since its long waveforms make it difficult to derive an exact ToA (which is the reason UWB is used in the first place). The attacker might exploit this tolerance in order to handle NB like the RIFs: instead of sending an advanced NB, they can stretch and replay the one sent by the genuine transmitter. Then, the attacker transmits all the following fragments as if the NB had been advanced, which results in a shorter pause between the NB and the first RSF. The resulting message should still be processed correctly by the attacked receiver and result in a successful attack. In the worst case, the attacker would have to reduce the drift $\delta_{c}$ to be within the accepted margins.

Second, long symbols can be advanced using the ED/LC attack~\cite{DBLP:conf/esas/ClulowHKM06}. Without knowing the symbol, an attacker could send an arbitrary waveform to trigger an earlier reception event in the receiver. As soon as learn the symbol from the actual NB fragment, they can use their power advantage to set the advanced NB symbol to the correct value.

\subsection{Analysis of S\&A}
\label{appendix:analysis}

\subsubsection{$\mathcal{T}(n_{att})$ for multiple RIFs}

For message formats with arbitrary RIFs, we can write: 
\begin{equation}
\label{eq:tnatt_4ab}
\mathcal{T}(n_{att}) =  \underbrace{\lfloor n_{att} / n_{RIF} \rfloor \cdot 1\si{ms}}_{\text{entire RIFs}} + \underbrace{(n_{att}\Mod{n_{RIF}})\cdot T_{bit}}_{\text{remaining bits}}
\end{equation}
Here, $n_{att}$ is the number of unpredictable bits, $n_{RIF}$ the number of bits in a RIF, and $T_{bit}$ the duration of a single bit. The two contributions are a result of the pauses between the RIFs: as long as $n_{att}$ is less than $n_{RIF}$, it grows linearly with $n_{att}$. However, for $n_{att}\geq_{RIF}$ the attacker accepts that an entire RIF becomes unpredictable, which shifts $t_{replay}$ (see \autoref{fig:v2_attack} to the end of the corresponding fragment. In this case, the attacker may just as well increase $\delta_{RIF}$ further and move $t_{replay}$ to the start of the \textit{next} RIF, since the pause between them does not contain any bits. These "free" advancements dominate $\delta_{RIF}$, since the duration of a RIF ($\approx 16\si{\micro\second}$) is small compared to the pause ($\approx 1\si{ms}$).

\subsubsection{Countermeasure for multiple RIFs}
We assume an attacker who aims at optimizing the probability $p_{succ}$  of achieving any distance reduction $\delta_{ToF}<0$. For $N_{RIF}\in\{2,4,8\}$, we have to use $\mathcal{T}$ as defined in \autoref{eq:tnatt_4ab}. If the condition on the clock drift consistency if met, the smallest sum $n_{att1} + n_{att2}$ that results in a distance reduction $\delta_{ToF}<0$ is $n_{tot}$, the total number of RIF bits in one ranging message: 

\begin{equation}
    \underset{n_{att1}+n_{att2}\in(0,2n_{tot})}{\arg\min}(\delta_{ToF}<0) = n_{tot}
\end{equation}

The maximum ToF reduction for such $n_{att1}$ and $n_{att2}$ is

\begin{equation}
\begin{aligned}
    \min_{n_{att1}+n_{att2}=n_{tot}}\delta_{ToF} &= \delta_{c}T_{pause}\\
\end{aligned}
\end{equation}

Where $T_{pause}\approx984\si{\micro\second}$ corresponds to the pause between fragments .
Because of $\mathcal{T}$, this result only holds when both $n_{atti}$ are an integer multiple of the bits in a single RIF, \textit{i.e}., $n_{att1}=x\cdot n_{RIF}, n_{att2}= (N_{RIF}-x)n_{RIF}$. The reason are the inter-RIF pauses: for these values, the attacker can exploit the trailing pause in ranging messages. This results in an overall reduction despite the enlargement caused by $\delta_{RSF}$ in \autoref{eq:cm_delta_tof2}. In contrast, any other $n_{att1}+n_{att2}=n_{tot}$ only compensate for the enlargement, resulting in $\delta_{ToF}=0$.

In the last step, we show that there is an $x$ for which $p_{suc}$ is maximal, which is the optimal solution we are interested in:
 \begin{equation}
 \begin{aligned}
    \underset{x\in\{1,..,N_{RIF}\}}{\arg\min}(p_{succ}=p_{RIF1}p_{RIF2}) = \frac{N_{RIF}}{2}\\
    \Rightarrow n_{att1}=n_{att2}=\frac{n_{tot}}{2}
    \end{aligned}
\end{equation}

Choosing any $n_{atti}>n_{tot}/2$ increases the ToF reduction, but at the cost of a lower probability. Therefore, $n_{atti}=n_{tot}/2$ maximizes the probability of any distance reduction. We conclude that setting $BER_{max}$ such that $p_{suc}$ is negligible prevents the S\&A attack.
\end{document}